\documentclass{JHEP3}
\usepackage{latexsym,amsmath,amssymb}
\pdfoutput=1
\usepackage{graphicx}
\usepackage{eufrak}
\usepackage{slashed}
\usepackage{bm}
\usepackage{epsfig}
\newcommand{\exclude}[1]{}
\usepackage{epsfig}
\usepackage{dcolumn}

\newcommand{\beq}{\begin{equation}}
\newcommand{\eeq}{\end{equation}}
\newcommand{\be}{\begin{eqnarray}}
\newcommand{\ee}{\end{eqnarray}}

\newcommand{\rar}{\rightarrow}

\def\dd{ \,\mathrm{d} }

\def\+{\dagger}
\def\la{\langle}
\def\ra{\rangle}
\def\<{\langle}
\def\>{\rangle}

\def\gmf{\gamma _{5}}
\newcommand{\Lqcd}{\Lambda_{\mathrm{QCD}}}

\title{${\cal P}$ and ${\cal CP}$ Violation and New Thermalization Scenario   in Heavy Ion Collisions  }

\author{  Ariel R. Zhitnitsky\\
Department of Physics \& Astronomy, University of British Columbia, Vancouver, B.C. V6T 1Z1, Canada}

\date{\today}

\abstract{ The violation of local ${\cal P}$ and ${\cal CP}$ invariance in QCD  has been a subject of intense discussions for the last couple of years as a result of very  interesting ongoing results coming from RHIC.
Separately, a new thermalization scenario for heavy ion collisions  through the event horizon as  a manifestation  of the Unruh effect,
has been also suggested. In this paper we argue that these two, naively unrelated phenomena, are actually two sides of the same coin as they  are deeply rooted into the same fundamental physics related to some very  nontrivial topological features of QCD.
We formulate  the universality conjecture for  ${\cal P}$ and ${\cal CP}$ odd effects in heavy ion collisions analogous to the universal thermal behaviour observed in all other  high energy  interactions. }

\begin{document}
\section{Introduction. Motivation}\label{motivation}
The main goal of this paper is to argue that two, naively unrelated, phenomena:\\
1.~local ${\cal{P}}$  and ${\cal{CP}}$ violation in QCD as studied  at RHIC\cite{Voloshin:2004vk,Selyuzhenkov:2005xa,Voloshin:2008jx,Abelev:2009uh,Abelev:2009tx}; and \\
2.~universal behaviour  of multihadron production described by  a universal hadronization temperature $T_H\sim (150- 200) ~{\text MeV}$\\ 
are in fact tightly related, as they describe 
different sides of the same fundamental physics. 

Before we present our arguments suggesting the common nature of these two phenomena, we review each effect separately as it is 
conventionally treated  today. Our next step is to take a fresh look at these effects and present some arguments suggesting that both these phenomena are in fact originated from the same fundamental physics and both are related to the very deep and nontrivial topological features  of QCD.
  \subsection{Local ${\cal{P}}$  and ${\cal{CP}}$ violation in QCD.  Charge separation effect}\label{P}
  The charge separation effect  \cite{Kharzeev:2004ey,Kharzeev:2007tn} can be explained in the following simple way. Let us assume that an  effective   $\theta (\vec{x}, t)_{ind}\neq 0$  is induced as a result of some non- equlibrium dynamics as suggested in refs.
 \cite{Kharzeev:1998kz,Kharzeev:1999cz,Halperin:1998gx,Fugleberg:1998kk,Buckley:1999mv,Buckley:2000aa}. The  $\theta (\vec{x}, t)_{ind}$ parameter enters the effective lagrangian as follows,  ${\cal L_{\theta}}=-\theta q$ where $ q \equiv \frac{g^2}{64\pi^2} \epsilon_{\mu\nu\rho\sigma} G^{a\mu\nu} G^{a\rho\sigma}$ such that local ${\cal{P}}$  and ${\cal{CP}}$
  invariance of QCD is broken on the scales where correlated $\theta (\vec{x}, t)_{ind}\neq 0$  is induced. As a result of this violation, one should expect a number of ${\cal{P}}$  and ${\cal{CP}}$ violating effects taking place in the region where $\theta (\vec{x}, t)_{ind}\neq 0$.  In particular, one should expect the separation of electric charge along the axis of magnetic field $\vec{B}$ or along the angular momentum $\vec{l}$
  if they are present in the region with $\theta (\vec{x}, t)_{ind}\neq 0$. 
  
  This area of research  became a very active  field in recent years mainly due to very interesting ongoing  experiments~\cite{Voloshin:2004vk,Selyuzhenkov:2005xa,Voloshin:2008jx,Abelev:2009uh,Abelev:2009tx}.    There is a number of different manifestations of this  local ${\cal{P}}$  and ${\cal{CP}}$  violation, see  \cite{Kharzeev:2007jp,Fukushima:2008xe,Kharzeev:2009pj,Kharzeev:2009fn,Buividovich:2009wi,Abramczyk:2009gb,Skokov:2009qp,Fukushima:2009ft,Fukushima:2010vw} and many additional references therein. In particular, in the presence of an external magnetic field $\vec{B}$ or in case of the rotating  system  with angular velocity $\vec{\Omega}$ there will be induced electric   current directed along $\vec{B}$ or $\vec{\Omega}$ correspondingly, resulting in   separation
 of charges along those directions as mentioned above. One can interpret the same effects as a generation of induced electric field $\vec{E}$ directed along $\vec{B}$ or $\vec{\Omega}$  resulting in corresponding electric current flowing along $\vec{J}\sim \vec{B}$ or 
 $\vec{J}\sim \vec{\Omega} $   directions. All these phenomena are  obviously   ${\cal{P}}$  and ${\cal{CP}}$
 odd effects. 
Non-dissipating, induced vector current  density    has the form:
  \be
  \label{J}
  \vec{J}=(\mu_L-\mu_R)\frac{e\vec{B}}{2\pi^2},
  \ee
  where ${\cal{P}}$ odd effect is explicitly present in this expression as the difference of chemical potentials of the right $\mu_R$ and left $\mu_L$ handed 
  fermions is assumed to be nonzero, $(\mu_L-\mu_R)\neq 0$. The combination $(\mu_L-\mu_R)$ can be thought
  as $\dot{\theta} (t) $ after  a corresponding $U(1)_A$ chiral time-dependent rotation is performed, see also~\cite{Kharzeev:2009fn}
  for a physical interpretation of the relation $(\mu_L-\mu_R)=\dot{\theta}$.
  
  Originally, formula (\ref{J}) has been derived in \cite{alekseev}, though in condensed matter context.  In QCD context formula (\ref{J})
  has been used in applications to neutron star physics where magnetic field is known to be large, and the corresponding $(\mu_L-\mu_R)\neq 0$ can be  generated   in neutron star environment as a result of  continuos ${\cal{P}}$  violating processes happening in nuclear matter\cite{Charbonneau:2007db,Charbonneau:2009ax}. It has been also applied to heavy ion collisions where an effective $(\mu_L-\mu_R)\neq 0$ 
  is locally induced. The effect  was estimated using the sphaleron transitions generating the topological charge density in the QCD plasma \cite{Kharzeev:2007jp,Fukushima:2008xe}. The effect was  coined as ``chiral magnetic effect" (CME)~\cite{Kharzeev:2007jp,Fukushima:2008xe}.  Formula (\ref{J}) has been also derived a numerous number of times using numerous variety of  techniques such as: effective lagrangian approach  developed in \cite{Son:2004tq}; explicit computation  approach
   developed in \cite{Metlitski:2005pr}; direct lattice computations \cite{Buividovich:2009wi,Abramczyk:2009gb}. 
   In addition, the effect has been studied in holographic models of QCD~\cite{ads/cft,Rubakov:2010qi,Brits:2010pw}. While there is a number of subtitles 
   in holographic description of the effect~\cite{Rubakov:2010qi,Brits:2010pw}, 
   it is fair to say: on the theoretical side the effect is a well established phenomenon. It remains to be seen if this phenomenon
   is   related in anyway to what has been actually experimentally observed ~\cite{Voloshin:2004vk,Selyuzhenkov:2005xa,Voloshin:2008jx,Abelev:2009uh,Abelev:2009tx}.  
   
  \subsection{Universal hadronization temperature $T_H\sim (150- 200) ~{\text MeV}$}\label{TT}
  Naively unrelated story goes as follows. 
  We start from the following general observation: over the years, hadron production studies in a variety of high energy collision experiments have shown a remarkably universal feature, indicating a universal hadronization temperature $T_H\sim (150- 200) ~{\text MeV}$.
  From $e^+e^-$  annihilation to $pp$ and $p\bar{p}$  interactions and further to collisions of heavy nuclei, with energies from a few GeV up to the TeV range, the production pattern always shows striking thermal aspects, connected to an apparently quite universal temperature around $T_H\sim (150- 200) ~{\text MeV}$ \cite{hagedorn}. 
  While  experimentally it  is well established phenomenon, it is very difficult to understand its nature as number of incident particles in $e^+e^-$  annihilation as well as in $pp$ and $p\bar{p}$  interactions is not sufficient even to talk about statistical averages. This observation motivated a number of  early attempts \cite{T_early} to interpret the resulting spectrum of particles as the Unruh radiation~\cite{Unruh:1976db,Unruh:1983ms,Birrell:1982ix} when the event horizon emerges as a result of strong interactions. The modern, QCD based formulation of this idea has been developed quite recently in refs.\cite{Kharzeev:2005iz,Kharzeev:2005qg,Kharzeev:2006zm,Kharzeev:2006aj,Castorina:2007eb,Satz:2008zza,Becattini:2008tx,Castorina:2008gf}. See also alternative approaches~\cite{Steinberg:2004vy,Evans:2007sf,Hatta:2008qx} 
leading to the same  ``apparently  thermal"   aspects of the produced particles.
  
  The key ingredient of the approach suggested in refs.~\cite{Kharzeev:2005iz,Kharzeev:2005qg,Kharzeev:2006zm,Kharzeev:2006aj,Castorina:2007eb,Satz:2008zza,Becattini:2008tx,Castorina:2008gf} is as follows: an observer moving with an acceleration $a$ experiences the influence of a thermal bath with an effective temperature 
  \beq
  \label{T}
  T=\frac{a}{2\pi}
  \eeq 
  which is conventional Unruh effect~\cite{Unruh:1976db}. The corresponding acceleration parameter $a$ in QCD is determined by the so-called ``saturation scale" $Q_s$ ~\cite{McLerran:1993ni,McLerran:1993ka} as suggested in refs.~\cite{Kharzeev:2005iz,Kharzeev:2005qg,Kharzeev:2006zm,Kharzeev:2006aj}, or by the string tension $\sigma$ as advocated in refs.~\cite{Castorina:2007eb,Satz:2008zza,Becattini:2008tx,Castorina:2008gf},  
  \beq
  \label{string}
 a\simeq Q_s ~~~~~{\text{or}} ~~~~~  a\simeq \sqrt{ 2\pi\sigma}.
  \eeq
  The   problem of calculating of the effective acceleration $``a"$ is obviously very hard problem of strongly interacting QCD. This problem of computation $``a"$  is not addressed in the present paper. We simply assume that such a description exists, in which case 
 the  relation (\ref{string})  explains the puzzle on why the temperature $T$ given by eq. (\ref{T}) is so universal, as it  it almost  independent  on type of processes and  the energy
  of  colliding particles, as  it is entirely determined by the fundamental $\Lambda_{QCD}$ scale. This ``apparent thermalization" originates from the event horizon in an accelerating frame: the incident hadron decelerates in an external colour field, which causes the emergence of the causal horizon. Quantum tunnelling  through this event horizon then produces a thermal final state of partons, in complete analogy with the thermal character of quantum Unruh radiation~\cite{Unruh:1976db,Unruh:1983ms,Birrell:1982ix}. One should emphasize that the Planck spectrum in this approach is not  resulted   from the  kinetics when the thermal equilibrium with temperature $T$ given by eq. (\ref{T}) is reached  due to the large number of collisions. Rather, the Planck  spectrum in high energy collisions is resulted from the stochastic tunnelling  processes when no information transfer occurs. In such circumstances   the spectrum must be thermal. Such interpretation would naturally explain another puzzle with the thermal spectrum in $e^+e^-, ~pp$ and $p\bar{p}$ high energy collisions when the statistical thermalization could  never be reached in those systems. 
    
  We adopt this viewpoint,
  and we have nothing new to add to the computations presented in refs.~\cite{Kharzeev:2005iz,Kharzeev:2005qg,Kharzeev:2006zm,Kharzeev:2006aj,Castorina:2007eb,Satz:2008zza,Becattini:2008tx,Castorina:2008gf}  in strongly coupled QCD. 
However, we interpret that the Planck spectrum observed  in high energy collisions somewhat differently
in comparison with papers mentioned above. Namely, we interpret the observed spectrum as the result of complete reconstruction of the QCD vacuum state in accelerating frame $``a"$, rather than due  to any specific properties of its excitations -- the partons.  Our interpretation does not change any previous  results  within this framework. However, the new interpretation will play a crucial role when we apply the same logic 
  and the same technique for discussions of  local ${\cal{P}}$  and ${\cal{CP}}$ violation in QCD, 
  which is the main subject of the present paper.

  $\bullet$ Our original contribution   is the study of some  specific topological fluctuations, 
 which we believe  are   responsible for  local ${\cal{P}}$  and ${\cal{CP}}$ violating processes observed at RHIC. 
  Those   vacuum topological fluctuations 
  will   be changed along with many other vacuum fluctuations  in accelerating frame $``a"$. 
 These    changes of the   ground state due to the acceleration as we shall see are describable  in terms of   the Veneziano ghost\footnote{not to be confused with conventional Fadeev Popov ghosts which appear in covariant quantization of non-abelian gauge theories}   which solves the $U(1)_A$ problem in QCD~\cite{ven,vendiv,witten,sch}. This  key  degree of freedom  (the Veneziano ghost)  has $\eta'-$ quantum numbers, and plays the role  similar to $\theta$ parameter  from section \ref{P}. Its  $0^{-+}$ quantum numbers, as we shall see,   play  the crucial role in  linking  two naively unrelated problems
  outlined  in two sections above: local ${\cal{P}}$  and ${\cal{CP}}$ violation in QCD, section \ref{P}, and universal hadronization temperature, section \ref{TT}.
  
  The paper is organized as follows.  In section \ref{Rindler} we  discuss the nature of  universality of hadronization temperature
  observed in numerous high energy experiments. 
  We adopt  the basic logic and philosophy  of refs.\cite{Kharzeev:2005iz,Kharzeev:2005qg,Kharzeev:2006zm,Kharzeev:2006aj,Castorina:2007eb,Satz:2008zza,Becattini:2008tx,Castorina:2008gf}.      We shall  explicitly demonstrate
  the observed thermal spectrum with temperature (\ref{T}) can be interpreted as   the direct consequence of the basic features of  known  Bogolubov's coefficients  in the accelerating frame
for a system moving with effective acceleration $``a"$. The temperature will be universal for all types of produced particles: massive or massless, charged or neutral, scalars, spinors, or vectors. Such a universality is similar to universal features of the Unruh radiation~\cite{Unruh:1976db,Unruh:1983ms,Birrell:1982ix}. 
   This result  will be our basic explanation for the thermal spectrum observed      in $e^+e^-, ~pp$ and $p\bar{p}$ high energy collisions when the statistical thermalization (due to the conventional collisions)  can not be ever  reached in those systems. 
   
 In  section \ref{Minkowski} we review  the resolution of the $U(1)_A$ problem and structure of the $\theta$ vacua ~\cite{ven,vendiv,witten,sch} by constructing the corresponding effective lagrangian.  We pay special attention to the structure of the Veneziano ghost, its contribution to the topological susceptibility with a ``wrong sign" (which is a key element in resolution of the $U(1)_A$ problem), 
    the unitarity, anomalous Ward Identities  and other important  properties of QCD. We formulate the physics of topological fluctuations (described by the Veneziano ghost in our framework)  in such a way that the relevant technique  can be easily generalized for the case of  accelerating frame. 
    
   Section \ref{Rindler1}  is devoted  to analysis of the Veneziano ghost in the accelerating frame.
    We shall  explicitly compute 
  the  Bogolubov's coefficients  to demonstrate   that the Veneziano  ghost contribution to energy  (being identically zero in Minkowski space)
  does not vanish anymore in accelerating frame, in 
  the Rindler space.      Therefore we identify the local ${\cal{P}}$  and ${\cal{CP}}$ violation in QCD  
   as a result of fluctuations of the vacuum $0^{-+}$ ghost field in the  accelerating frame.   
   We conclude with section \ref{applications} where we list the  possible tests      discriminating this framework  from the previously suggested mechanisms.   
   The readers interested in   applications only 
   may skip sections \ref{Rindler},\ref{Minkowski},\ref{Rindler1} and immediately jump  to section \ref{applications}.
   \exclude{
   In this section   we also present an  analogy with cosmic microwave background (CMB)  analysis where
   the ${\cal{P}}$ violating fluctuations on scales of size of visible universe $\sim $ 10 Gyr  (in contrast with  $\sim\text{fm} $ scale which is the subject of the present work)  apparently   have been observed. We argue that the fundamental nature of  ${\cal{P}}$ violating fluctuations seen in CMB sky
   is the same as ${\cal{P}}$    violating correlations observed at RHIC ~\cite{Voloshin:2004vk,Selyuzhenkov:2005xa,Voloshin:2008jx,Abelev:2009uh,Abelev:2009tx} though   at drastically different scales: $\sim $  Gpc  versus $\sim\text{~fm} $. 
}    
     \section{Universal hadronization temperature   as  the Unruh effect}\label{Rindler}
     As we mentioned above, we adopt  the basic logic and philosophy  of refs.\cite{Kharzeev:2005iz,Kharzeev:2005qg,Kharzeev:2006zm,Kharzeev:2006aj,Castorina:2007eb,Satz:2008zza,Becattini:2008tx,Castorina:2008gf}. Essentially, 
  the previously developed picture can be explained  in a simplified way  as follows: the incident parton decelerates in an external colour field.    The causal horizon  emerges  as a result of this strong interaction. Quantum tunnelling  through the emergent event horizon then produces a thermal final state of partons. The hard part of the problem, the computation 
  of the acceleration $``a"$ is not addressed in the present work. The acceleration  must be an universal number (which we assume to be the case), not sensitive to a colour representation of the incident particles. Once the universal acceleration is reached, the produced particles will automatically have a thermal spectrum.
  
  Our interpretation of this physics is somewhat different: instead of tunnelling  of real particles we rather speak  about  changes of the ground state. Initially, the ground state   was   a pure quantum state in Minkowski space.  In accelerating frame  the Rindler  observers  (who do not ever have access to the entire space-time as a result of emerging horizon)  would see the same ground state as a mixed state (rather than pure state) filled by particles with Planck  spectrum. 
 This  new interpretation will be quite important  when we apply the  
  same technique in section \ref{Rindler1} for discussions of the topological vacuum fluctuations in accelerating frame 
  because there will be no real asymptotic states which would correspond to those topological vacuum fluctuations. Indeed, as we argue below, the Veneziano ghost does not contribute to absorptive parts
  of any correlation functions, but only to the real parts. 
 According to our logic these topological vacuum fluctuations in accelerating frame  will serve as a  ${\cal{P}}$  and ${\cal{CP}}$ odd background for physical  fluctuations of quarks and gluons. As we argue below,  these topological vacuum fluctuations will be eventually  responsible for the   ${\cal{P}}$  and ${\cal{CP}}$ violating correlations observed at RHIC. 
    
   As we mentioned above, the computation  of the acceleration parameter $``a"$   is  a hard problem  of strongly interacting gauge theory which  is not addressed in the present work.  
  Instead, we adopt the entire framework   and 
    treat the acceleration $``a"$    as a free parameter of the theory. This parameter, in principle, could have had any value. In nature  $``a"$  is not really a free parameter, but expressed in terms of $\Lambda_{QCD}$
as eq. (\ref{string}) states. However, to simplify things, we will be working in the limit $1~ {\text{GeV}}\gg a\gg m_q$ where our arguments  can be made precise, though numerically $a\simeq 1~ {\text{GeV}}$ as estimate (\ref{string}) shows. Also: we have nothing new to add to the previous arguments \cite{Kharzeev:2005iz,Kharzeev:2005qg,Kharzeev:2006zm,Kharzeev:2006aj,Castorina:2007eb,Satz:2008zza,Becattini:2008tx,Castorina:2008gf} suggesting that the acceleration (\ref{string}) and the corresponding temperature (\ref{T}) are universal for different processes. Rather we take a given  accelerating parameter assuming $a\ll  1~ {\text{GeV}}$ and   study   the changes in the ground state (in comparison with Minkowski vacuum state) which occur due to the acceleration. In  our analysis in what follows we consider the radiation  of a single massless scalar field to  demonstrate   the most important features of the accelerating frame; generalization to vector and spinor fields in the Rindler space is also known, but shall not be discussed in the present work. 

\subsection{Rindler space}

   We follow notations   \cite{Birrell:1982ix} in our analysis
   and separate  the space time into four quadrants $F$ (future), $P$ (past), $ L$ (left wedge) and $R$ (right wedge).  We will choose the origin such that these regions are defined by $t>|x|$, $t<-|x|$, $x<-|t|$ and $x>|t|$ respectively.  While no single region contains a Cauchy surface, the union of the left and right regions $L$ and $R$ plus the origin does contain many Cauchy surfaces, for example $t=0$.   
We will write the Minkowski metric with the sign convention
\beq
ds^2=dt^2-dx^2 -dy^2-dz^2,
\eeq
In the quadrant $R$, called  the right Rindler wedge, one may define the coordinates $(\xi^R,\eta^R, y, z)$ via the transformations
\beq
\label{eta}
t=\frac{e^{a\xi^R}}{a}{\text{sinh}}\ a\eta^R, ~~~~~  x= \frac{e^{a\xi^R}}{a}{\text{cosh}}\ a\eta^R
\eeq
where $a$ is a dimensional  constant. We may define coordinates $(\xi^L,\eta^L, y, z)$ in the left Rindler wedge $L$ in a similar way with the signs of both $t$ and $x$ reversed~\cite{Birrell:1982ix}.   
In these new coordinates the metric is presented as 
\beq
\label{metric}
ds^2=e^{2a\xi}(d\eta^2-d\xi^2)-dy^2-dz^2.
\eeq
Without loosing any generalities we ignore in what follows a trivial dependence on $y, z$ coordinates. 
It is important  to emphasize that the coordinates $(\eta^R, \xi^R) $ cover only a quadrant of Minkowski space,
namely the wedge $x>|t|$. Lines of constant $\xi$ are hyperbolae
\beq
x^2-t^2=a^{-2} e^{2a\xi} = \text{world lines}.
\eeq
They represent the world lines of uniformly accelerated observers with proper acceleration given by
\beq
\label{a}
 a e^{-a\xi} = \text{proper acceleration}.
\eeq
Thus, lines of large positive $\xi$ (far from $x=t=0$) represent weakly accelerated observers, while large negative $\xi$ correspond a high proper acceleration. The observer's proper time $\tau$ 
is
\beq
\label{tau}
\tau= \eta e^{a\xi} = \text{observer's proper time}.
\eeq
Uniformly accelerated observers will be referred to as Rindler observers. 
It is important to emphasize that  $L$ and $R$ regions are separated by the event horizons such that no events in $L$ can be witnessed in $R$ and vice versa.
In different words, regions $L$ and $R$ represent two causally disjoint universes. 
Left wedge $L$ with $x<|t|$ is obtained by changing the signs in eq. (\ref{eta}).
The sign reversals in $L$ mean that increasing $t$ corresponds to decreasing $\eta$ which implies that 
time-like Killing vector being $+\partial_{\eta}$ in R becomes   $-\partial_{\eta}$ in L   in contrast with Minkowski space 
where $\partial_t$ is time-like Killing vector in entire space. This feature plays an important  role in selection of positive frequency modes
in $L$ and $R$ wedges as discussed below. 

In terms of these variables the picture of high energy collision (in very simplified way)  can be represented as follows
(detail derivations and explanations are presented below in next subsection). Two energetic particles 
approach the interaction region $(x=t=0)$ along the light cone from $x=t=-\infty$ and $x=-t=\infty$. When the colliding particles
have non-vanishing space-like transverse momenta $k^2=-k_{\perp}^2$ their world lines are located off the light cone as described in \cite{Kharzeev:2005iz}.
Due to the interaction
the initial particles  experience acceleration (\ref{a}) which we assume to be  a constant $``a"$ to simplify  our computations.
In this case one can switch to the Rindler frame where the question is formulated as follows: how is the initial ground state expressed as a Fock state in the Rindler frame? As we shall see below it will not be a ground state any more in the accelerating frame. Rather it will be a superposition of excited states which include both: the $L$ and $R$ components separated by the horizon. 
We shall see that the corresponding   excitations   have the thermal spectrum as the direct consequence 
  of the  Bogolubov's coefficients' properties. 
It is a different way to explain 
the same physics which was described in refs.\cite{Kharzeev:2005iz,Kharzeev:2005qg,Kharzeev:2006zm,Kharzeev:2006aj,Castorina:2007eb,Satz:2008zza,Becattini:2008tx,Castorina:2008gf} as a tunnelling through the event horizon.

\subsection{Quantum Fields  in Rindler space} 
As we mentioned above, the dynamics along $y,z$ directions is trivial, and we ignore it in what follows to simplify the notations.
The wave equation of a free massless field $\phi (t, \vec{x})$ possesses standard orthonormal mode solutions  
\beq
\label{M}
u_k=\frac{1}{\sqrt{4\pi\omega}} e^{-i\omega t+i k x}.
\eeq
 such that   we can expand it in terms of  complete orthonormal basis $u_k (t, \vec{x})$  
\be
\label{expansion-M}
\phi (t, \vec{x})=\sum_{k}\left[ b_ku_k(t,\vec{x})+b_k^{\dagger}u_k^*(t,\vec{x})\right].
\ee
 The   commutation relations take   the form,
\be\label{comm2}
\left[b_k, b_{k'}\right]&=&0 \, , \; [b_k^{\dagger}, b_{k'}^{\dagger}]=0 \, , \; [b_k, b_{k'}^{\dagger}]=\delta_{kk'} \, ,
\ee
while the ground state in Minkowski space $|0\>_M$ is defined as usual
\be
\label{vacuum}
  b_k|0_M\>=0 \, , ~~~ \forall k \, .
\ee
The number operator $\mathrm{N}$ and  the Hamiltonian $\mathrm{H}$ for $\phi$  field in these notations have the standard  form
\be
\label{M-N}
\mathrm{N}=\sum_k b_k^{\dagger}b_k \, ,~~~
\mathrm{H}=\sum_k \omega_k b_k^{\dagger}b_k    \, ,
\ee
where normal ordering    is implied (but not explicitly shown) in all formulae presented below, such that
 the ground state satisfies the standard conditions  
\be
\label{H=0}
\< 0_M|\mathrm{H} |0_M\>=0 \, ,
 ~~~\<0_M| \mathrm{N} |0_M\>=0 \, .
\ee
  
  We want to describe the ground state defined by eq. (\ref{vacuum}) in terms of the Rindler coordinates. For the metric (\ref{metric})
  the corresponding modes are:
\be
\label{R}
^Ru_k= \frac{1}{\sqrt{4\pi\omega}} e^{ik\xi^R-i\omega\eta^R}~~~~ {\rm in~ R} ,~~~ ^Ru_k=0~~~~~ {\rm in~ L}  
\ee
\be
\label{L}
^Lu_k=\frac{1}{\sqrt{4\pi\omega}} e^{ik\xi^L+i\omega\eta^L} ~~~~~~{\rm in~ L} , ~~~^Lu_k=0 ~~~~~~~{\rm in~ R} 
\ee
The set ({\ref{R}) is complete in region R, while (\ref{L}) is complete in L, but neither is complete in on all of Minkowski space. 
However, both sets together are complete. 
The sign difference corresponds to the fact that  a right moving wave in R moves towards increasing value of $\xi$, while in L it moves toward decreasing value of $\xi$.  In any case, these modes are positive frequency modes with respect to the time-like Killing vector $+\partial_{\eta}$ in R and  $-\partial_{\eta}$ in L.   The fact that (\ref{R}) and (\ref{L}) have the same functional form as (\ref{M})
is a consequence of the conformal triviality of the system.
 Thus the Rindler modes (\ref{R}) and (\ref{L})  represent a good basis for quantizing the $\phi$ field, as good as the Minkowski basis (\ref{M}).
Therefore,  one can use modes (\ref{R}) and (\ref{L}) to expand the field $\phi$
\be
\label{expansion-R}
 \phi=\sum_k\frac{1}{\sqrt{4\pi\omega}}(b^L_ke^{ik\xi^L+i\omega \eta^L}+b^{L\dagger}_ke^{-ik\xi^L-i\omega \eta^L}+b^R_ke^{ik\xi^R-i\omega \eta^R}+b^{R\dagger}_ke^{-ik\xi^R+i\omega \eta^R}),
\ee
where $ b_k^L, b_k^R$ satisfy the following commutation relations,
\be\label{comm_2}
\left[b_k^{R}, b_{k'}^{R}\right]&=&0 \, , \; [b_k^{{R}\dagger}, b_{k'}^{{R}\dagger}]=0 \, , \; [b_k^{R}, b_{k'}^{{R}\dagger}]=\delta_{kk'} \, ,
\ee
and similar for $L-$ Rindler wedge operators $ b_k^L$. 
The Rindler vacuum state is defined in terms of these operators as follows, 
\be
\label{vacuum-R}
   b_k^R|0_R\>=0 \, , ~~~ \forall k \, .
\ee

We need to compute  the corresponding Bogolubov's coefficients which relate two alternative expansions 
(\ref{expansion-M}) and (\ref{expansion-R})  in order to answer the question: how is the initial ground state expressed as a Fock state in the Rindler frame?
The simplest way to compute these coefficients is to note that although $^Ru_k $ and $^Lu_k$ are not analytic, the two combinations 
\be
\label{analytic}
 \exp{(\frac{\pi\omega}{2a})} ~ ^Ru_k  + \exp{(-\frac{\pi\omega}{2a})}~   ^Lu_{-k}^*  \\ \nonumber
  \exp{(-\frac{\pi\omega}{2a})}~   ^Ru_{-k}^*   + \exp{(\frac{\pi\omega}{2a})}~ ^Lu_k 
 \ee
 are analytic and bounded\cite{Unruh:1976db}.  These modes share the positivity frequency analyticity properties of the Minkowski modes (\ref{M}), therefore, they must also share a common vacuum state, see below precise definition. Therefore, instead of expansion (\ref{expansion-M}) with modes (\ref{M}) we can expand $\phi$  in terms of (\ref{analytic}) as
\be 
\label{expansion-M2}
\phi&=&\sum_k\frac{1}{\sqrt{ 4\pi\omega}}  \cdot \frac{1}{\sqrt{(e^{\pi\omega/a}-e^{-\pi\omega/a})}} \Big[b^1_k(e^{\frac{\pi\omega}{2a}+ik\xi^R-i\omega\eta^R}+e^{\frac{-\pi\omega}{2a}+ik\xi^L-i\omega\eta^L})  \nonumber \\
&+&b^2_k(e^{\frac{\pi\omega}{2a}+ik\xi^L+i\omega\eta^L}+e^{\frac{-\pi\omega}{2a}+ik\xi^R+i\omega\eta^R})  \nonumber \\
&+&b^{1\dagger}_k(e^{\frac{\pi\omega}{2a}-ik\xi^R+i\omega\eta^R}+e^{\frac{-\pi\omega}{2a}-ik\xi^L+i\omega\eta^L})\nonumber\\
&+&b^{2\dagger}_k(e^{\frac{\pi\omega}{2a}-ik\xi^L-i\omega\eta^L}+e^{\frac{-\pi\omega}{2a}-ik\xi^R-i\omega\eta^R})\Big], 
\ee
where $ b_k^1, b_k^2$ satisfy the following commutation relations,
\be\label{comm_M2}
\left[b_k^{(1,2)}, b_{k'}^{(1,2)}\right]&=&0 \, , \; [b_k^{{(1,2)}\dagger}, b_{k'}^{{(1,2)}\dagger}]=0 \, , \; [b_k^{(1,2)}, b_{k'}^{{(1,2)}\dagger}]=\delta_{kk'} \, .
\ee
 The
 Minkowski vacuum state is determined in terms of these operators as
 \be
\label{vacuum-M}
  b_k^1|0_M\>=0 \, , ~~~  b_k^2|0_M\>=0 \, , ~~~ \forall k \, .
\ee
This equation   replaces eq. (\ref{vacuum}) as it defines the same ground state $|0_M\>$ because 
  both sets    share a common vacuum state as explained above.
  Matching coefficients in (\ref{expansion-R})  with  (\ref{expansion-M2})  one finds the Bogoliubov's coefficients~~\cite{Unruh:1976db,Birrell:1982ix},
\be
\label{Bogolubov}
b^L_k=\frac{e^{-\pi\omega/2a}b^{1\dagger}_{-k}+e^{\pi\omega/2a}b^2_k}{\sqrt{e^{\pi\omega/a}-e^{-\pi\omega/a}}}~~~~~~
b^R_k=\frac{e^{-\pi\omega/2a}b^{2\dagger}_{-k}+e^{\pi\omega/2a}b^1_k}{\sqrt{e^{\pi\omega/a}-e^{-\pi\omega/a}}}.  
\ee
Now consider an accelerating Rindler observer at $\xi= $const. As we mentioned above, such an observer's proper time is proportional to $\eta$, see eq. (\ref{tau}). The vacuum for this observer is determined by (\ref{vacuum-R}) as this is the state associated with the positive frequency modes with respect to $\eta$.
A Rindler observer in $R$  will measure the energy using  the Hamiltonian $\mathrm{H}^{R}$ and the number operator  $\mathrm{N}^{R}$ which are given by
\be
\label{H-RR}
\mathrm{N}^{R}=\sum_k  b_k^{R\dagger}b_k^{R}  \, ,~~~~~~ \mathrm{H}^{R}=\sum_k \omega_k b_k^{R\dagger}b_k^{R}  \, .
\ee
Similar expressions are also valid for a $L$-Rindler observer.
The  Hamiltonian $\mathrm{H}^{R}$ and the number operator  $\mathrm{N}^{R}$  in the R-Rindler accelerating frame
 have the same form as for conventional Minkowski expressions  (\ref{M-N}). However, they are expressed in terms of the  
 different operators which select a different ground state $|0_R\>$ as defined by eq. (\ref{vacuum-R}).
 It is obvious  that the ground state in R-wedge, $|0_R\>$ satisfies the standard conditions, 
\beq
\label{N=0}
  \< 0_R|{\mathrm{H}}^{R}|0_R\> =0 ,~~~   \< 0_R|{\mathrm{N}}^{R}|0_R\> =0   ,
  \eeq
as it should.

However, if the initial system is prepared as  the Minkowski vacuum state $ |0_M\> $ defined by (\ref{vacuum-M})
(or what is the same (\ref{vacuum})) a Rindler observer using the same expression for the number operator
(\ref{H-RR}) will observe the following number of particles  in mode $k$, 
\be
\label{Planck}
\< 0_M |   {\mathrm{N}}^{R} |0_M\>=  \< 0_M |   b_k^{R\dagger}b_k^{R}  |0_M\>= 
  \frac{  e^{-\pi\omega/a}}{{(e^{\pi\omega/a}-e^{-\pi\omega/a})}}= \frac{1 }{(e^{2\pi\omega/a}-1)}, 
  \ee
  where we used the Bogolubov's coefficients (\ref{Bogolubov}) to express $b_k^{R}  $ in terms of $b_k^{(1,2)}$.

This  is the central formula of this section and represents nothing but the 
 the conventional Unruh effect~\cite{Unruh:1976db,Birrell:1982ix}.
In context of high energy collisions the Planck spectrum given by eq. (\ref{Planck})  
  was described  in refs. \cite{Kharzeev:2005iz,Kharzeev:2005qg,Kharzeev:2006zm,Kharzeev:2006aj,Castorina:2007eb,Satz:2008zza,Becattini:2008tx,Castorina:2008gf} as a tunnelling through the event horizon.  In our description   the same
  physical effect is resulted from  restriction of Minkowski  vacuum $ \left| 0_M \right>$ to a single  Rindler wedge region 
 where it becomes a thermal   state (rather than a pure quantum state) with temperature $T=\frac{a}{2\pi}$.
This structure is precisely  resulted   from expression of the Minkowski ground state  $ \left| 0_M \right>$
in terms of the excited states in $L$ and $R$ regions  when  the combination  $b_k^{R\dagger}b_{-k}^{L\dagger} $
which  includes  the  operators from    causally disconnected regions $L$ and $R$,
  enters formula  for $ \left| 0_M \right>$ as can be seen 
  from explicit construction:
 \be
 \label{squeezed}
 |0_M\>= \prod_k\frac{1}{ \sqrt{(1-e^{-2\pi\omega/a})}}\exp\left[   e^{-\pi\omega/a}b_k^{R\dagger}b_{-k}^{L\dagger}
 \right]   \left| 0^{R}\right>  \otimes \left| 0^{L}\right>, 
 \ee
 where we take into account that the operators in the $L, R$ basis correspond to the decompositions with support in only one wedge such that the right hand side is represented by the tensor product $ \left| 0^{R}\right>  \otimes \left| 0^{L}\right>$.
 The crucial point in this  relation   is the fact that  the operators from  different causally disconnected regions $L$ and $R$ enter the same expression (\ref{squeezed}), and therefore, there is a correlation between causally disconnected regions $L$ and $R$.  However, 
 as discussed in ref.\cite{Unruh:1976db}, 
 one can not use these correlations   to send signals. 

 In context of high energy collisions the picture advocated in refs.  \cite{Kharzeev:2005iz,Kharzeev:2005qg,Kharzeev:2006zm,Kharzeev:2006aj,Castorina:2007eb,Satz:2008zza,Becattini:2008tx,Castorina:2008gf} when two causally disconnected regions are connected as a result of the  tunnelling through the event horizon,  manifests itself  in eq. (\ref{squeezed}) by emergence of   the  combination $b_k^{R\dagger}b_{-k}^{L\dagger} $ when $L$ and $R$ components are  simultaneously present  in   eq. (\ref{squeezed}). The expression (\ref{squeezed})  also shows that the Planck spectrum observed in high energy collisions can be interpreted as a result of {\it preparation} of the ground state $ |0_M\>$ even before  collision 
 develops. This interpretation naturally explains the puzzle with rapid ``thermalization" observed in all high energy collisions.
  Many other observed consequences (such as dependence on transverse momenta $k_{\perp}^2$, saturation scale at low energies and/or peripheral AA collisions, dependence on strange mass quark, and many others) also find their simple explanations in  this framework. They have been discussed in those references in great details,  and we have nothing new to 
  add to those discussions as the basic logic of our  approach and the one advocated in refs.  \cite{Kharzeev:2005iz,Kharzeev:2005qg,Kharzeev:2006zm,Kharzeev:2006aj,Castorina:2007eb,Satz:2008zza,Becattini:2008tx,Castorina:2008gf} is the same as  both pictures naturally  lead to the Planck spectrum (\ref{Planck}). 
  
   Still, we want to add one more comment which my shed some light on mysterious  ``apparent thermalization" effect resulted from the acceleration.  
   The ``apparent thermalization" can be understood as entanglement type behaviour when 
   the Planck spectrum    emerges as a result  of the description    in terms of the density matrix in $R$ region by ``tracing out" over the degrees of freedom associated with 
 inaccessible states in $L$-region. In different words, 
  the Bogolubov transformations (\ref{Bogolubov}) 
  describe a  construction  when a total system is divided into two subsystems with the horizon separating them.  This is the deep physics reason why  the Planck spectrum (\ref{Planck})  emerges for a subsystem. 
 It is known that a number of nontrivial physics effects (including the entanglement)
  are described by the common surface separating such two sub- systems, i.e. by the horizon separating $L$ and $R$ Rindler wedges.   The particle production in    the framework advocated in refs.  \cite{Kharzeev:2005iz,Kharzeev:2005qg,Kharzeev:2006zm,Kharzeev:2006aj,Castorina:2007eb,Satz:2008zza,Becattini:2008tx,Castorina:2008gf} occur exactly from this horizon region, while in our framework it can be interpreted   as a result of entanglement. One should also note that typical quark and gluon  vacuum fluctuations develop in this accelerating environment characterized by temperature (\ref{T}), not in laboratory frame, and  not in center of mass frame. Once acceleration ceases to exist, a   detector  in laboratory frame will measure a particle  distribution according to the temperature (\ref{T}) as the acceleration ends almost instantaneously, and produced particles do not have time
  to make any adjustments to a new environment. In this respect,  it is very similar to measurements of CMB (cosmic micro wave background) radiation as the CMB photons 
  being produced at the last scattering at temperature $T\simeq 2.7~$K nevertheless do not change their properties during the next $\sim$ 14 billion years.

$\bullet$   To conclude this section: we have not derived anything new which was not previously known. However, we presented the ``new thermalization" scenario  advocated in refs.  \cite{Kharzeev:2005iz,Kharzeev:2005qg,Kharzeev:2006zm,Kharzeev:2006aj,Castorina:2007eb,Satz:2008zza,Becattini:2008tx,Castorina:2008gf} as a result of reconstruction of the vacuum state in accelerating system. This new interpretation will be a crucial element in section \ref{Rindler1} when we argue that some very specific  topological fluctuations 
   are responsible for violation of local ${\cal P}$ and ${\cal CP}$ invariance in QCD observed at RHIC. We would not be able to use the quasi-classical technique   developed in refs.  \cite{Kharzeev:2005iz,Kharzeev:2005qg,Kharzeev:2006zm,Kharzeev:2006aj,Castorina:2007eb,Satz:2008zza,Becattini:2008tx,Castorina:2008gf} to discuss  corresponding fluctuations as no classical trajectories of real particles propagating in external classical colour fields  exist for these type of  fluctuations. This is because the relevant topological    vacuum fluctuations, as we discuss below,
    are not related to any absorptive parts of any physical correlation functions. Rather,  the topological fluctuations which will be  main subject of this paper may contribute only to the real parts of the correlation functions (such as topological susceptibility) see next section.   In Minkowski space similar contributions normally treated as the subtraction constants. In the present case of the accelerating frame, the corresponding ``subtraction constant" becomes a ``subtraction function" which depends on acceleration and which is sensitive to the global properties of the space-time. 
As we shall see below,  the approach developed in the present section  is well suited to attack this problem as everything in section  \ref{Rindler1} will be  formulated precisely in appropriate terms of vacuum fluctuations.

   \section{The $\theta$ vacua, $U(1)_A$ effective lagrangian and the Veneziano ghost in  Minkowski spacetime}\label{Minkowski} 
The main goal of this section is to single out (identify)  the  fields which describe the topological fluctuations in QCD.
It turns out that these relevant fields (in a specific gauge) can be represented as (pseudo)scalar colour-singlet fields such that one can immediately apply the technique developed 
in previous section \ref{Rindler} to describe the  corresponding vacuum fluctuations in  accelerating frame.  As we shall argue below in section \ref{Rindler1}  those vacuum topological fluctuations in accelerating frame may be responsible for    ${\cal{P}}$ and ${\cal{CP}}$ violation  observed at RHIC.  It is quite obvious that
those configurations must be related to $\theta$ dependence, topological charge density,  and other related problems. 

Therefore,  we start this section by reviewing  the $\theta$ dependence in QCD and   the standard resolution of $U(1)_A$ problem~\cite{ven,vendiv,witten,sch}. We follow~\cite{UZ} to identify the
 relevant for the present work degrees of freedom  (Veneziano ghost) which saturates topological correlation functions.
After  integrating   the ghost  out (as was done in the original paper~\cite{vendiv}) one reproduces the $\eta'$ mass, which was the main result of~\cite{vendiv}.  We   keep the Veneziano ghost explicitly
  as it can not be integrated out in accelerating frame. Moreover,   it will play a central  role in our following discussions when we consider the accelerating frame   relevant for description of high energy collisions\exclude{ in which case the cancellation is not quite complete.}. As we shall argue in section \ref{Rindler1} 
the corresponding topological fluctuations  in accelerating frame might be the pivotal  source of ${\cal{P}}$ and ${\cal{CP}}$ violating fluctuations  observed at RHIC.  
\exclude{  It turns out that the obtained Lagrangian includes, along with Veneziano ghost,  another paired massless companion  field such that in 4d it \emph{identically coincides}  with the corresponding expression for the 2d Schwinger model~\cite{KS}.  Unitarity in this new description is explicit due  to the appearance of the ghost's companion which exactly cancels all unphysical contributions in complete analogy  with the Gubta-Bleuler~\cite{G,B} conditions in QED
when unphysical photon's polarizations cancel each other in physical Hilbert space. 
 }

\subsection{The Lagrangian and the ghost}\label{lagr}

The starting point for our analysis will be the Lagrangian   as proposed  in~\cite{vendiv,sch}   in large $N_c\gg 1$ limit whose general form reads
\be\label{lag}
{\cal L} &=& {\cal L}_0+\frac{1}{2}\partial_\mu \eta' \partial^\mu \eta'  + \frac{N_c}{b f_{\eta'}^2}q^2 - \left(\theta-\frac{\eta'}{f_{\eta'}} \right)q \nonumber 
+ N_f m_q |\<\bar{q}q\>| \cos\left[ \frac{\eta'}{f_{\eta'}} \right]  + g.f. \, , \nonumber \\
q &=& \frac{g^2}{64\pi^2} \epsilon_{\mu\nu\rho\sigma} G^{a\mu\nu} G^{a\rho\sigma} \equiv \frac{1}{4} \epsilon_{\mu\nu\rho\sigma} G^{\mu\nu\rho\sigma} \, ,\\
G_{\mu\nu\rho\sigma} &\equiv& \partial_\mu A_{\nu\rho\sigma} - \partial_\sigma A_{\mu\nu\rho} + \partial_\rho A_{\sigma\mu\nu} - \partial_\nu A_{\rho\sigma\mu} \, , \label{four} \nonumber  \\
A_{\nu\rho\sigma} &\equiv& \frac{g^2}{96\pi^2} \left[ A_\nu^a \stackrel{\leftrightarrow}{\partial}_\rho A_\sigma^a - A_\rho^a \stackrel{\leftrightarrow}{\partial}_\nu A_\sigma^a - A_\nu^a \stackrel{\leftrightarrow}{\partial}_\sigma A_\rho^a + 2 gC_{abc} A_\nu^a A_\rho^b A_\sigma^c \right] \, . \nonumber 
\ee
where we explicitly keep only relevant for the present work  degrees of freedom, such as $\eta'$ and the topological density $q$, while all others (including $\pi, K, \eta$) are assumed to be in ${\cal L}_0$, and shall not be mentioned in this paper.   In this Lagrangian $g.f.$ means gauge fixing term for three-form $A_{\mu\nu\rho}$, and the coefficient $b\sim m_{\eta'}^2$ is fixed by the Witten-Veneziano relation for the topological susceptibility in pure gluodynamics.   The fields $A_\mu^a$ are the usual $N_c^2-1$ gauge potentials for chiral QCD and $C_{abc}$ the $SU(N_c)$ structure constants.  The constant $f_{\eta'}\simeq f_{\pi}$ is the $\eta'$ decay constant, while $m_q$ is the quark mass, and $ \<\bar{q}q\>$ is the chiral condensate.

The three-form $A_{\nu\rho\sigma}$ is an abelian totally antisymmetric gauge field which, under colour gauge transformations $\Lambda^a$ behaves as
\be\label{transf}
A_{\nu\rho\sigma} \rar A_{\nu\rho\sigma} + \partial_\nu \Lambda_{\rho\sigma} - \partial_\rho \Lambda_{\nu\sigma} - \partial_\sigma \Lambda_{\rho\nu} \, ,  ~~
\Lambda_{\rho\sigma} \propto A_\rho^a \partial_\sigma \Lambda^a - A_\sigma^a \partial_\rho \Lambda^a \, ,
\ee
such that the four-form $G_{\mu\nu\rho\sigma}$ is gauge invariant as it should.  The term proportional to $\theta$ is the usual $\theta$-term of QCD and appears in conjunction with the $\eta'$ field in the correct combination as dictated by the Ward Identities (WI).  The constant $b$ is a \emph{positive} constant which  determines   the topological susceptibility in pure gauge theory,
\be
\label{top}
i\int \!\dd^4x \la T\{q(x), q(0)\}\ra_{YM}=-\frac{f_{\pi}^2}{8N_c} b \, .
\ee
The constant $b$ saturates  the topological susceptibility
 and has a wrong sign (in comparison with contribution from any real physical states),  the property which motivated the term ``Veneziano ghost''.  
 Indeed, a physical state  of mass $m_G$, momentum $k\rightarrow 0$  and coupling $\la 0| q| G\ra= c_G$ contributes to the topological susceptibility  with the sign which is opposite to (\ref{top}), 
 \be
\label{G}
i \lim_{k\rightarrow 0} \int \!\dd^4x e^{ikx} \la T\{q(x), q(0)\}\ra  \sim i ~\lim_{k\rightarrow 0}  \la  0 |q|G\ra \frac{i}{k^2-m_G^2}\la G| q| 0\ra \simeq \frac{|c_G|^2}{m_G^2} \geq 0.
\ee
 However, the positive sign for $b$ (and negative sign for the topological susceptibility (\ref{top})) is what is required to extract the physical mass for the $\eta'$ meson, $m_{\eta'}^2 \simeq b/2N_c \neq 0$, see the original reference~\cite{vendiv} for a thorough discussion.
 
  One can interpret the field $A_{\mu\nu\rho}$ as a collective mode  of the original gluon fields, which in the infrared leads to a pole in the unphysical subspace and provides  a finite contribution with a wrong sign to the topological susceptibility (\ref{top}).  We know about the existence of this very special degree of freedom and its properties from the resolution of the famous $U(1)_A$ problem: integrating out the $q$ field (as shown below) provides the mass for the $\eta'$ meson.
Of course $b=0$ to any order in perturbation theory because $q(x)$ is a total divergence $q = \partial_\mu K^\mu $.  However, as we learnt from~\cite{ven,witten}, $b\neq 0$ due to the non-perturbative infrared physics; in fact, in the chiral limit $m_{\eta'}^2 \sim b$.

One can integrate out the scalar field $q$   since there is no kinetic term associated to it.  This is indeed the procedure followed by Di~Vecchia and Veneziano in their original paper~\cite{vendiv}, the outcome of which, as follows from~(\ref{lag}),  is
\be\label{lag0}
{\cal L} = {\cal L}_0 + \frac{1}{2}\partial_\mu \eta' \partial^\mu \eta'  -\frac{bf_{\eta'}^2}{4N_c}\left(\theta -\frac{\eta'}{f_{\eta'}}\right)^2  + N_f m_q |\<\bar{q}q\>| \cos\left[ \frac{\eta'}{f_{\eta'}} \right] \, ,
\ee
where all the dependence on the three-form $A_{\nu\rho\sigma}$ has disappeared.  This formula explicitly shows that $\eta'$ receives a non-vanishing mass in the chiral limit, $m_{\eta'}^2 \simeq b/2N_c \neq 0$ due to the non-zero magnitude of the coefficient $b$, which enters~(\ref{lag}) and~(\ref{top}).  This formula also reproduces the notorious Witten-Veneziano relation for the topological susceptibility in pure gluodynamics if one substitutes $b=2N_c m_{\eta'}^2$ into the expression~(\ref{top}) for the topological susceptibility.

As we mentioned above, we want to keep the ghost   hidden in $A_{\mu\nu\rho}$ explicitly. We shall now choose the Lorenz-like gauge 
\be
\label{gauge}
(\partial^{\rho}A_{\mu\nu\rho}) \simeq \left( \partial_\mu K_\nu - \partial_\nu K_\mu \right)=0 \, , ~~~~~~~
K_\mu \equiv \epsilon_{\mu\nu\rho\sigma} A^{\nu\rho\sigma} \, , ~~~~~ q = \partial_\mu K^\mu \, ,
\ee
in which we will carry out our manipulations.  It is the same gauge which was discussed in the original paper~\cite{vendiv}.   We choose to work only with the longitudinal part of the $K_{\mu}$ field because only this longitudinal part determines the topological density $q = \partial_\mu K^\mu$, and eventually leads to a  non-vanishing contribution to the topological susceptibility~(\ref{top}).  Therefore, we  write the longitudinal part of $K_\mu $ as
\be\label{Phidef}
K_\mu \equiv \partial_\mu \Phi \, ,
\ee
such that the expression for the topological density takes the form
\be\label{boxdef}
q = \partial_\mu K^\mu = \Box \Phi \, ,
\ee
where $\Phi$ is a new scalar field of mass dimension 2. We notice that the gauge condition~(\ref{gauge}) is automatically satisfied with our definition~(\ref{Phidef}).  Now our Lagrangian~(\ref{lag}) can be expressed in terms of the $\Phi$ field as follows
\be\label{lag1}
{\cal L} &=& {\cal L}_0 + \frac{1}{2}\partial_\mu \eta' \partial^\mu \eta'  +N_f m_q |\<\bar{q}q\>| \cos\left[ \frac{\eta'}{f_{\eta'}} \right] \\ \nonumber
 &+& \frac{1}{2m_{\eta'}^2f_{\eta'}^2} \Phi \Box \Box \Phi + \left(\frac{\eta'}{f_{\eta'}}\right) \Box \Phi  - \theta \Box \Phi \, ,
\ee
where we plugged in the coefficient $b\rar 2N_c m_{\eta'}^2$ as the Witten-Veneziano relation requires.  If we integrate out the $ \Box \Phi $ field we return to the expression~(\ref{lag0}) which describes the physical massive $\eta'$ field alone.

As usual, the presence of 4-th order operator  $\Phi \Box \Box \Phi$ is a signal that the ghost is present in the system and may be quite dangerous. However, we know from the original form~(\ref{lag}) that the system is unitary, well defined etc, in different words, it does not present any problem associated with the ghost.  
\exclude{
It is convenient to define a new field $ \phi_2$ which is a combination of the the original $\eta'$ field and $\Phi$ as
\be\label{phi2def}
\eta' \equiv \left( \phi_2 +\frac{\Phi}{f_{\eta'}} \right) \, ,
\ee
which serves to complete the squares in~(\ref{lag1}) in such a way that one can eliminate the term  $\int \!\dd^4x \eta' \Box \Phi=- \int \!\dd^4x \partial_\mu \Phi \partial^\mu \eta'$.  The Lagrangian now takes the form
\be\label{lag2}
{\cal L} &=& \frac{1}{2} \partial_\mu \phi_2 \partial^\mu \phi_2 +N_f m_q |\<\bar{q}q\>| \cos\left[ \frac{\phi_2}{f_{\eta'}} +\frac{\Phi}{f_{\eta'}^2}\right]
\\ \nonumber
&+& \frac{1}{2m_{\eta'}^2f_{\eta'}^2} \Phi \left[ m_{\eta'}^2\Box +  \Box \Box \right] \Phi \, .
\ee
The next step is to   explicitly represent $ \left[ m_{\eta'}^2\Box +  \Box \Box \right] $ operator as the combination of the ghost
and a massive physical $\eta'$ using the standard trick by 
  writing  the inverse operator as follows
\be
\label{inverse}
\frac{m_{\eta'}^2}{ \Box \Box + m_{\eta'}^2\Box  }= \left(\frac{1}{ - \Box - m_{\eta'}^2  } -  \frac{1}{ - \Box  } \right) \, .
\ee
In analogy with the 2d Kogut and Susskind (KS) model~\cite{KS}, we will call  the massive scalar field as $\hat\phi$  (which is nothing but the physical  massive $\eta'$ meson) while the massless ghost is $\phi_1$. 
}
 One can redefine all  fields in   such a way that 
the final Lagrangian can  be  expressed as follows\cite{UZ}:
\be\label{lagKS}
{\cal L} &=& \frac{1}{2} \partial_\mu \hat\phi \partial^\mu \hat\phi + \frac{1}{2} \partial_\mu \phi_2 \partial^\mu \phi_2 - \frac{1}{2} \partial_\mu \phi_1 \partial^\mu \phi_1 \\
&-& \frac{1}{2} m_{\eta'}^2 \hat\phi^2 + N_f m_q |\<\bar{q}q\>| \cos\left[ \frac{\hat\phi + \phi_2 - \phi_1}{f_{\eta'}} \right] \nonumber\, ,
\ee
where all fields have now canonical dimension one in four dimensions.
We claim that the Lagrangian~(\ref{lagKS}) is that part of QCD which describes long distance physics in our context.   Notice that~(\ref{lagKS}) is exactly \emph{identical} to that proposed by Kogut and Susskind (KS) in~\cite{KS} for the 2d Schwinger model, see also \cite{toy} in given context.   The unitarity and other important properties of QFT are satisfied in our 4d system~(\ref{lagKS}) in the same way as they are satisfied in ~\cite{KS} for the 2d Schwinger model
as will be reviewed in the next subsection \ref{GB}.   
The Veneziano ghost in QCD is represented in our   notations by the $\phi_1$ field in~(\ref{lagKS}) and it is always accompanied by its companion, the massless field $\phi_2$.  These two fields cancel each other in every gauge invariant matrix element once the 
 auxiliary (similar to Gupta-Bleuler~\cite{G,B}) conditions on the physical Hilbert space are imposed, see below.
 \exclude{
 Yet, there is one place where the ghost has physical consequences, and that is the mass spectrum which, through the Witten-Veneziano mass formula~\cite{ven,witten} relates the mass of the $\eta'$ to the topological susceptibility of the model.  It is precisely the topological susceptibility which enjoys the uncancelled contribution of the ghost $\phi_1$ with its companion $\phi_2$  because the topological density operator $q$ is expressed in terms of $\Phi$, see~(\ref{boxdef}). At the same time, the Green's function for $\Phi$ field   is a combination of the Green's functions for the ghost $\phi_1$ and  massive  $\hat\phi$  field (which is nothing but physical $\eta'$ in conventional notations), but not of the companion $\phi_2$ as one can see from eq.  (\ref{inverse}).    It is this property that leads to a nonzero contribution to the topological susceptibility from the Veneziano ghost $\phi_1$ with the ``wrong" sign, which  eventually leads to the $\eta'$ mass.
 }
 
 Using the explicit expression for the Green's function and the expression for the topological density 
 $q = \partial_\mu K^\mu = \Box \Phi \, $ after simple algebraic manipulations one can  represent the topological susceptibility for  QCD
 in model (\ref{lag})  in the chiral limit $m_q\rightarrow 0$
 in the following way, 
 \be
\label{top1}
\chi_{QCD}\equiv i\int \!\dd^4x \la T\{q(x), q(0)\}\ra_{QCD}= -\frac{f_{\eta'}^2 m_{\eta'}^2}{4} \cdot \int d^4x\left[ \delta^4 (x)- m_{\eta'}^2 D^c (m_{\eta'}x)\right]
\ee
where $D^c (m_{\eta'}x)$ is the Green's function of a free massive particle with standard normalization $\int d^4x m_{\eta'}^2 D^c (m_{\eta'}x)=1$. In this expression the $\delta^4(x)$ represents the ghost contribution with the required ``wrong"  sign while  the term proportional  to $ D^c (m_{\eta'}x)$ represents the $\eta'$ contribution. The ghost's contribution can be also thought as the Witten's contact term not related to any propagating degrees of freedom. The topological susceptibility $\chi_{QCD} (m_q= 0)=0 $ vanishes in the chiral limit  as a result of exact cancellation of two terms entering (\ref{top1}) in complete accordance with WI.   When $m_q\neq 0$ the cancellation is not complete and $\chi_{QCD}\simeq m_q \<\bar{q}q\>$.

\begin{figure}[t]
\begin{center}
 \includegraphics[width = 0.5\textwidth]{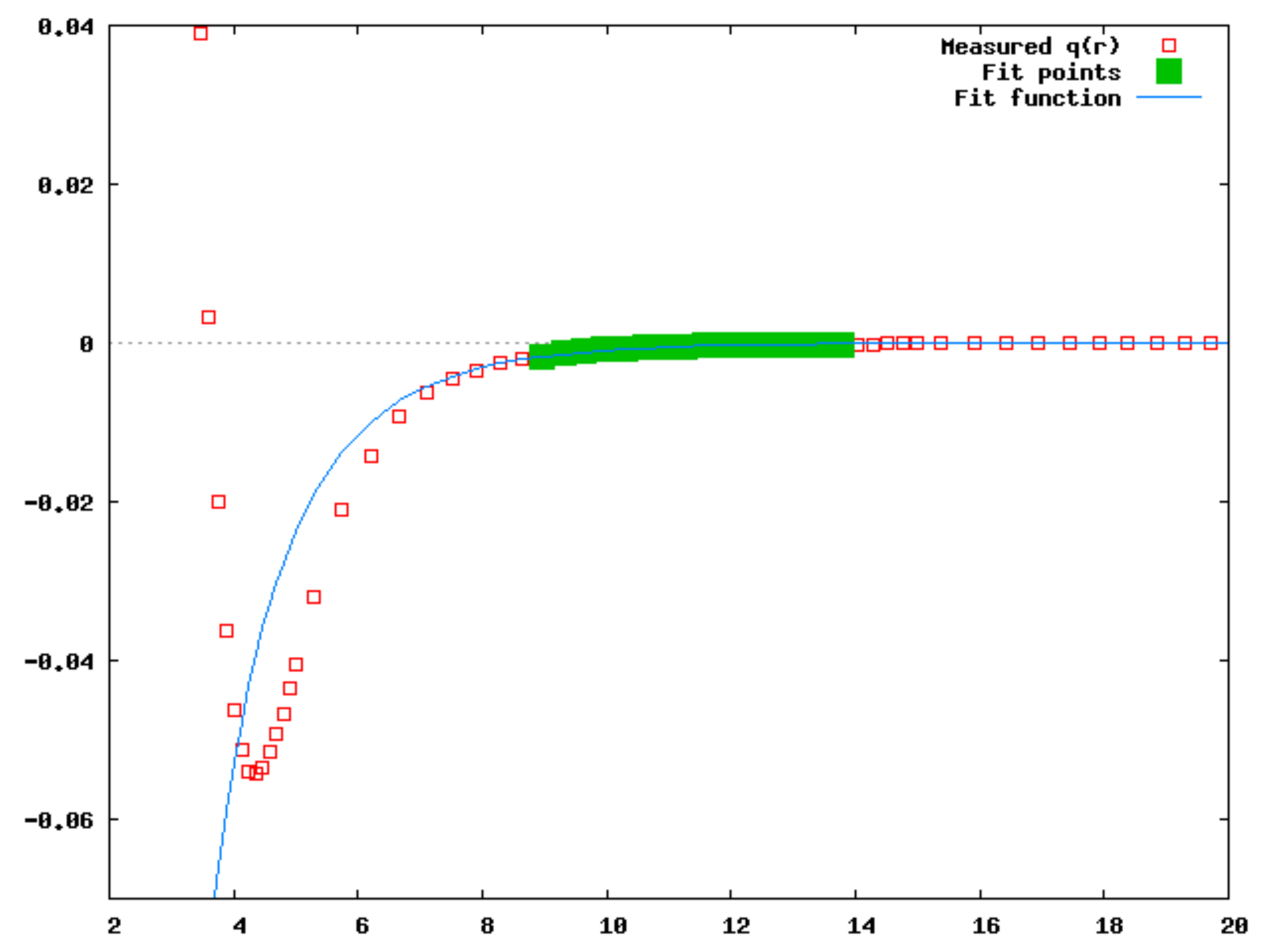}
 \caption{\label{chi-lattice}
The density of the topological susceptibility $\chi(r)\sim \< q(r), q(0)\>$ as function of separation $r$ 
such that $\chi_{QCD}\equiv \int d r \chi(r)$, adapted  from~\cite{lattice}. Plot explicitly shows the presence 
of the contact term with the ``wrong sign" (narrow peak around $r\simeq 0$) represented by the Veneziano ghost in our framework.  
 }
\end{center}
\end{figure}

One should emphasize that the presence of the contact term (described by the Veneziano ghost in our framework) $\sim  \delta^4 (x)$ in eq. (\ref{top1}) is well established and well understood phenomenon in QCD. In particular, it has been studied on the lattice, see e.g.
\cite{lattice} and references therein\footnote{ A warning signal with the signs: the physical degrees of freedom in  Euclidean space (where the lattice computations are performed)  contribute  to topological susceptibility $\chi_{QCD}$ with the negative sign, while the contact term (the Veneziano ghost) contributes with the positive sign, in contrast with our discussions in Minkowski space.} where a narrow peak around $r\simeq 0$ and a smooth behaviour in  extended region  of  $r\sim \text{fm}$  with the opposite signs   have been  seen as a result of numerical computations. We reproduce Fig.\ref{chi-lattice} for illustration purposes adapted from ref.\cite{lattice} where these crucial elements are explicitly present on the plot. As we mentioned previously, 
the main goal of the present paper is to study precisely this (unphysical and  non-propagating)  effective degree of freedom leading to  $\sim  \delta^4 (x)$ in eq. (\ref{top1}) (and  represented by a narrow peak at $r\sim 0$ on Fig. \ref{chi-lattice}) 
when we   accelerate our  system.  As we discuss in great details in  \cite{Zhitnitsky:2010ji} the ghost
does not contribute to absorptive parts of any correlation functions (after all, it is not an asymptotic degree of freedom). However, it does contribute to the real part as plot  on Fig.\ref{chi-lattice}  explicitly shows. 
 
 As we shall argue below,   the  topological nature of the  ghost and its    $0^{-+}$ quantum numbers  may  play a crucial role in understanding of local ${\cal{P}}$  and ${\cal{CP}}$ violation in QCD
   observed at RHIC~\cite{Voloshin:2004vk,Selyuzhenkov:2005xa,Voloshin:2008jx,Abelev:2009uh,Abelev:2009tx}.
   Before we proceed with our description of the Veneziano ghost in accelerating frame we would like to demonstrate  that 
   the Veneziano ghost is harmless (e.g. it does not violate unitarity)  in spite of its negative sign  in  the Lagrangian (\ref{lagKS}).

\subsection{Unitarity and the ghost}\label{GB}
We follow Kogut and Susskind  construction ~\cite{KS} in order to demonstrate the unitarity of our system when 
the Veneziano ghost $\phi_1$ and its partner $\phi_2$ explicitly present in the Lagrangian (\ref{lagKS}). 
\exclude{
As in the KS model, we      impose the following canonical equal-time commutation relations for the fields $\hat\phi$, $\phi_1$, and $\phi_2$  \be\label{comm}
\left[ \hat\phi (t, \vec{x})\, , \, \partial_t \hat\phi (t, \vec{y})\right] &=& i \delta^3 (\vec{x}-\vec{y})\, , \nonumber\\
\left[ \phi_1 (t, \vec{x})\, , \, \partial_t \phi_1 (t, \vec{y})\right] &=&- i \delta^3 (\vec{x}-\vec{y})\, , \\
\left[ \phi_2 (t, \vec{x})\, , \, \partial_t \phi_2 (t, \vec{y})\right] &=& i \delta^3 (\vec{x}-\vec{y})\, , \nonumber
\ee
whence we evince that $\phi_1$ is a massless ghost field, and its propagator will have an opposite  sign in comparison with the conventional fields which is reflected in eq. (\ref{inverse}) for the Green's function of the ghost. }
The cosine interaction term (\ref{lagKS}) includes vertices between the ghost and the other two scalar fields, but it can in fact be shown
in complete analogy with ~\cite{KS} that, once appropriate auxiliary (similar to Gupta-Bleuler~\cite{G,B}) conditions on the physical Hilbert space are imposed, the unphysical degrees of freedom $\phi_1$ and $\phi_2$ drop out of every gauge-invariant matrix element, leaving the theory well defined, i.e., unitary and without negative normed physical states, just as in the Lorentz invariant quantization of electromagnetism.  Specifically, this is achieved by demanding that the {\it positive frequency part } of the free massless combination $(\phi_2 - \phi_1)$ annihilates the physical Hilbert space:
\be\label{gb}
(\phi_2 - \phi_1)^{(+)} \left|{\cal H}_{\mathrm{phys}}\right> = 0 \, .
\ee
With this additional requirement the quantum theory built on the Lagrangian~(\ref{lagKS}) is well defined in any respect, and the physical sector of the theory  exactly coincides with~(\ref{lag0}) which was obtained by a trivial integrating out procedure~\cite{vendiv}.

In particular, one can explicitly  check that the expectation value for any physical state in fact vanishes as a result of the subsidiary condition~(\ref{gb}):
\be
\label{H=0M}
\< {\cal H}_{\mathrm{phys}}| \mathrm{H} |{\cal H}_{\mathrm{phys}}\>=0 \, .
\ee
In different words, all these ``dangerous'' states which can produce arbitrary negative energy do not belong to the physical subspace defined by eq.~(\ref{gb}).  
\exclude{The same argument applies to the operator $\mathrm{N}$ with identical result
\be
\label{N=0M}
\< {\cal H}_{\mathrm{phys}}| \mathrm{N} |{\cal H}_{\mathrm{phys}}\>=0 \, ,
\ee
where we can see explicitly the pairing and cancelling mechanism at work. 
}
 Therefore, the main conclusion is that the description in terms of the ghost is equivalent to well-known standard procedure of integrating out the ghost  field leading to well-known expression (\ref{lag0}) for the effective low energy lagrangian.

 In the next section we want to study the dynamics of the ghost fields in the accelerating frame. The   corresponding technique for a non-interacting  (pseudo)scalar field has been previously developed and presented in section \ref{Rindler}. In what follows we consider the chiral limit $m_q\rightarrow 0$
 such that the lagrangian describing the ghost field and its partner  (\ref{lagKS}) is precisely represented by the combination of two massless non- interacting fields\footnote{ The fluctuations of the physical massive $\eta'$ field can be obviously neglected. It will be ignored in what follows.}
 in which case the corresponding Bogolubov's coefficients have been previously computed (\ref{Bogolubov}). Numerically, the chiral limit in fact implies that the acceleration parameter $a\gg m_q$. In addition, as we mentioned previously, we want to keep $``a"$ as a  free parameter of the theory in spite of the fact that
 in nature it is fixed by the string tension (\ref{string}). Therefore, in all discussions below we consider the following hierarchy of scales 
 \beq
 \label{hierarchy}
 1~ {\text{GeV}} \gg a\gg m_q
 \eeq 
 in order  to separate the effects
 topological fluctuations due to the acceleration $``a"$ from conventional QCD fluctuations with typical scales $\sim  {\text{GeV}} $.
  
  \section{ ${\cal{P}}$  and ${\cal{CP}}$ violating fluctuations  in accelerating frame}\label{Rindler1}
  Our goal here is to understand the behaviour of the system (\ref{lagKS}) in the chiral limit $m_q=0$ in accelerating frame. 
  These fields have quantum numbers $0^{-+}$ and their long wave fluctuations may produce observable ${\cal{P}}$  and ${\cal{CP}}$  odd effects as discussed below. 
  There are many other  fluctuations, of course, in the system  due to conventional  quarks and gluons. 
  However, the ghost field $\phi_1$ and its partner $\phi_2$ are unique degrees of freedom in many respects, and their fluctuations, hopefully, can be separated from all other vacuum fluctuations. 
  
  The main point is as follows.  As we discussed above, the Bogolubov's coefficients (\ref{Bogolubov})  have the property that they are exponentially suppressed for $\omega\gg a$. Therefore, the typical wave lengths of fluctuations  related to acceleration are $\lambda\geq a^{-1}$. If $a\simeq   1~ {\text{GeV}}$ as estimation (\ref{string}) suggests,  it would be very difficult to disentangle  these fluctuations from conventional  fluctuations of  quarks and gluons
  with the same typical scale $\lambda\simeq  {\text{GeV}}^{-1}$. 
  However, we work in the limit $ a\ll 1~ {\text{GeV}}$ where such separation (at least theoretically) is a possibility\footnote{In fact, we argue in the next section \ref{applications} that in heavy ion collisions such conditions indeed could be achieved experimentally.}. In case  $ a\ll 1~ {\text{GeV}}$ all colour fields will still fluctuate with typical $\lambda\simeq  {\text{GeV}}^{-1}$ as a result of confinement
  which implies that  all fluctuations are effectively  gapped. It is in a drastic contrast with fluctuations of colourless  ghost field $\phi_1$ and its partner $\phi_2$. The ghost remains massless even when interactions are present, as its pole  (in unphysical Hilbert space) is topologically protected as discussed above
  in section \ref{Minkowski}. 
  Indeed, non vanishing contribution to the topological susceptibility (\ref{top}) constructed from the operators which are total derivatives 
  $q=\partial_{\mu}K^{\mu}$ may  only come from  (unphysical) massless pole. In different words, the  typical wave lengths   of $\phi_1$ and $\phi_2$ fields related to acceleration are $\lambda\geq a^{-1}$ for arbitrary small $a$.
 Based on this comment,  we ignore  in this section the conventional  fluctuations of quarks and gluons with $\lambda\simeq  {\text{GeV}}^{-1}$; we return to them in section \ref{applications}  when we produce  some numerical estimates by comparing the energetics of the ghost  fields $\phi_1, \phi_2$ and conventional quarks and gluons.

  \subsection{Veneziano ghost in the accelerating frame at $ 1~ {\text{GeV}} \gg a\gg m_q$
}
  As we mentioned above, in the region $ 1~ {\text{GeV}} \gg a\gg m_q$ the problem is  reduced to free massless fields $\phi_1$ and $\phi_2$ with
  GB like constraint. Therefore,  one can explicitly use the formalism developed earlier in section~\ref{Rindler}. In particular, one can compute
  the Bogolubov's coefficients   (\ref{Bogolubov}),  construct  the Hamiltonian and the number operator 
  for the ghost field $\phi_1$  and its partner $\phi_2$ in accelerating frame as it was done previously, see eq.(\ref{H-RR}).  The next step is to compute the corresponding expectation values when the system is being prepared as Minkowski vacuum $|0_M\>$ evolves in the accelerating background. Technically it is exactly the same    problem  of our previous   computations of  the Planck spectrum (\ref{Planck})  detected by a Rindler observer in a model of  a single  massless particle.  

We start with  expansion of   the ghost field $\phi_1$ and its partner     $\phi_2$ using the  modes (\ref{R}) and (\ref{L}) as we have done previously 
(\ref{expansion-R}),
\be
\label{expansion-RR}
\phi_1=\sum_k\frac{1}{\sqrt{4\pi\omega}}(a^L_ke^{ik\xi^L+i\omega \eta^L}+a^{L\dagger}_ke^{-ik\xi^L-i\omega \eta^L}+a^R_ke^{ik\xi^R-i\omega \eta^R}+a^{R\dagger}_ke^{-ik\xi^R+i\omega \eta^R}) \\ \nonumber
\phi_2=\sum_k\frac{1}{\sqrt{4\pi\omega}}(b^L_ke^{ik\xi^L+i\omega \eta^L}+b^{L\dagger}_ke^{-ik\xi^L-i\omega \eta^L}+b^R_ke^{ik\xi^R-i\omega \eta^R}+b^{R\dagger}_ke^{-ik\xi^R+i\omega \eta^R}).
\ee
The Rindler vacuum state is defined as usual,
\be
\label{vacuum-R1}
 a_k^R|0_R\>=0 \, ,  ~~~  b_k^R|0_R\>=0 \, , ~~~ \forall k \, .
\ee

In Minkowski space we can proceed exactly along the same line as we have done in section \ref{Rindler}. Namely,  instead of expansion  with modes (\ref{M}) we can expand $\phi_1$  and $\phi_2$  in terms of (\ref{analytic}) as follows:
\be 
\label{expansion-MM1}
\phi_1&=&\sum_k\frac{1}{\sqrt{ 4\pi\omega}}  \cdot \frac{1}{\sqrt{(e^{\pi\omega/a}-e^{-\pi\omega/a})}} \Big[a^1_k(e^{\frac{\pi\omega}{2a}+ik\xi^R-i\omega\eta^R}+e^{\frac{-\pi\omega}{2a}+ik\xi^L-i\omega\eta^L})  \nonumber \\
&+&a^2_k(e^{\frac{\pi\omega}{2a}+ik\xi^L+i\omega\eta^L}+e^{\frac{-\pi\omega}{2a}+ik\xi^R+i\omega\eta^R})  \nonumber \\
&+&a^{1\dagger}_k(e^{\frac{\pi\omega}{2a}-ik\xi^R+i\omega\eta^R}+e^{\frac{-\pi\omega}{2a}-ik\xi^L+i\omega\eta^L})\nonumber\\
&+&a^{2\dagger}_k(e^{\frac{\pi\omega}{2a}-ik\xi^L-i\omega\eta^L}+e^{\frac{-\pi\omega}{2a}-ik\xi^R-i\omega\eta^R})\Big], \nonumber\\
\phi_2&=&\sum_k\frac{1}{\sqrt{ 4\pi\omega}}  \cdot \frac{1}{\sqrt{(e^{\pi\omega/a}-e^{-\pi\omega/a})}} \Big[b^1_k(e^{\frac{\pi\omega}{2a}+ik\xi^R-i\omega\eta^R}+e^{\frac{-\pi\omega}{2a}+ik\xi^L-i\omega\eta^L})  \nonumber \\
&+&b^2_k(e^{\frac{\pi\omega}{2a}+ik\xi^L+i\omega\eta^L}+e^{\frac{-\pi\omega}{2a}+ik\xi^R+i\omega\eta^R})  \nonumber \\
&+&b^{1\dagger}_k(e^{\frac{\pi\omega}{2a}-ik\xi^R+i\omega\eta^R}+e^{\frac{-\pi\omega}{2a}-ik\xi^L+i\omega\eta^L})\nonumber\\
&+&b^{2\dagger}_k(e^{\frac{\pi\omega}{2a}-ik\xi^L-i\omega\eta^L}+e^{\frac{-\pi\omega}{2a}-ik\xi^R-i\omega\eta^R})\Big], 
\ee
where $ b_k^1, b_k^2$ satisfy the following commutation relations,
\be\label{comm_2R}
\left[b_k^{(1,2)}, b_{k'}^{(1,2)}\right]&=&0 \, , \; [b_k^{{(1,2)}\dagger}, b_{k'}^{{(1,2)}\dagger}]=0 \, , \; [b_k^{(1,2)}, b_{k'}^{{(1,2)}\dagger}]=\delta_{kk'} \, ,
\ee
 whereas $a_k^1, a_k^2$ for the ghost field $\phi_1$  satisfy
\be
\label{comm_1R}
\left[a_k^{(1,2)}, a_{k'}^{(1,2)}\right]&=&0 \, , \; [a_k^{{(1,2)}\dagger}, a_{k'}^{{(1,2)}\dagger}]=0 \, , \; [a_k^{(1,2)}, a_{k'}^{{(1,2)}\dagger}]=-\delta_{kk'} 
\ee
where again the sign minus appears in these commutation relations. The
 Minkowski vacuum state is determined as usual 
\be
\label{vacuum-new}
a_k^1|0\>=0 \, , ~~~a_k^2|0\>=0 \, , ~~~ b_k^1|0\>=0 \, , ~~~  b_k^2|0\>=0 \, , ~~~ \forall k \, .
\ee
  The Bogolubov's coefficients for $\phi_1$ and $\phi_2$ fields relating the description in Minkowski  and Rindler spaces can be computed exactly in the same way as it was done before, see eq. (\ref{Bogolubov}),
\be
\label{Bogolubov1}
a^L_k=\frac{e^{-\pi\omega/2a}a^{1\dagger}_{-k}+e^{\pi\omega/2a}a^2_k}{\sqrt{e^{\pi\omega/a}-e^{-\pi\omega/a}}}~~~~~~
a^R_k=\frac{e^{-\pi\omega/2a}a^{2\dagger}_{-k}+e^{\pi\omega/2a}a^1_k}{\sqrt{e^{\pi\omega/a}-e^{-\pi\omega/a}}}\\  
b^L_k=\frac{e^{-\pi\omega/2a}b^{1\dagger}_{-k}+e^{\pi\omega/2a}b^2_k}{\sqrt{e^{\pi\omega/a}-e^{-\pi\omega/a}}}~~~~~~
b^R_k=\frac{e^{-\pi\omega/2a}b^{2\dagger}_{-k}+e^{\pi\omega/2a}b^1_k}{\sqrt{e^{\pi\omega/a}-e^{-\pi\omega/a}}}. \nonumber
\ee
Now consider an accelerating Rindler observer at $\xi= $const. As we discussed previously, such an observer's proper time is proportional to $\eta$. The vacuum for this observer is determined by (\ref{vacuum-R1}) as this is the state associated with the positive frequency modes with respect to $\eta$.
A Rindler observer in R wedge will measure the energy and particle density using  the Hamiltonian $\mathrm{H}^{R}$ 
and density operator $\mathrm{N}^{R}$  which are given by
(a similar formula applies for L wedge as well),
\be
\label{H-R}
\mathrm{H}^{R}=\sum_k \omega_k\left(b_k^{R\dagger}b_k^{R}- a_k^{R\dagger}a_k^{R}\right) \, , ~~~
\mathrm{N}^{R}=\sum_k \left(b_k^{R\dagger}b_k^{R}- a_k^{R\dagger}a_k^{R}\right) \, .
\ee
The subsidiary condition~(\ref{gb}) defines the physical subspace for accelerating Rindler observer 
\be\label{gb-R}
\left(a_k^{R}-b_k^{R}\right) \left| {\cal H}_{\mathrm{phys}}^{R}\right> = 0 \, ,  
\ee
such that the exact cancellation between $\phi_1$ and $\phi_2$ fields holds  for any physical state defined by eq. (\ref{gb-R}), i.e.
\beq
\label{R=0}
  \left<{\cal H}_{\mathrm{phys}}^R|\mathrm{H}^{R}|{\cal H}_{\mathrm{phys}}^{R}\right> = 0 
  \eeq
as it should.

However, if the system is prepared as  the Minkowski vacuum state $ |0_M\> $ defined by eq.(\ref{vacuum-new}) a Rindler observer using the same expressions for the number operator and Hamiltonian 
(\ref{H-R}) will observe the following amount  of energy in mode $k$, 
\be
\label{RR1}
\< 0 | \omega_k\left(b_k^{R\dagger}b_k^{R}- a_k^{R\dagger}a_k^{R}\right)  |0\>= 
  \frac{2\omega e^{-\pi\omega/a}}{{(e^{\pi\omega/a}-e^{-\pi\omega/a})}}= \frac{2\omega }{(e^{2\pi\omega/a}-1)},
\ee
  where we used the Bogolubov's coefficients (\ref{Bogolubov1})  to express $a_k^{R}, b_k^{R}  $ in terms of $a_k^{(1,2)}, b_k^{(1,2)}$.
  This formula (up to a numerical coefficient) has been reproduced in ref.\cite{ohta} by using a different technique.
This  is the central result of this section and is a direct analog of  Planck spectrum given by eq. (\ref{Planck})  discussed previously
for the conventional massless particle with the only difference of factor 2 in front which is result of extra degeneracy: we have two degrees of freedom $\phi_1$ and $\phi_2$ instead of one scalar field $\phi$ from section \ref{Rindler}. The crucial point here is as follows. No cancellation between the ghost $\phi_1$ and its partner $\phi_2$ could occur in the expectation value~(\ref{RR1}), in net contrast with eq.~(\ref{R=0}). Technically, a   ``non-cancellation" of unphysical degrees of freedom (\ref{RR1}) in accelerating frame is   a result of opposite sign in commutator (\ref{comm_1R}) along with negative sign in Hamiltonian (\ref{H-R}). 

We will discuss this important  point in great details in the next subsection in non-technical, intuitive way.
However, we want to emphasize that this result (\ref{RR1}) 
should not be interpreted as actual emission of ghost modes, as they are not the asymptotic states in Minkowski spacetime  in the remote past and future, and therefore they can not propagate to infinity in contrast with conventional Unruh effect, see appendix A for details. Rather, one should interpret    (\ref{RR1})  as  an additional time dependent contribution to the vacuum energy in accelerating background in comparison with Minkowski space-time. This extra energy is  entirely ascribable to the presence of the unphysical (in Minkowski space) degrees of freedom. 
\exclude{In different words,
these unphysical degrees of freedom do not contribute to {\it absorptive parts} of the correlation functions. However, they do contribute to the {\it real parts} of the correlation functions. In Minkowski space such a contribution  is normally  represented by a ``subtraction constant".
In a curved/time-dependent background  this ``subtraction constant" becomes a ``subtraction function" sensitive to the entire space-time. }
We interpret the extra contribution to the energy observed by the Rindler observer as  a result of  formation of a specific   configuration,      the ``ghost condensate" ~\cite{Zhitnitsky:2010ji}, rather than a presence of ``free particles"   prepared in a specific mixed state which can be detected. This extra term should be treated as a result of very unique vacuum fluctuations, not related to any absorptive contributions. 
The observational effects of this extra vacuum contribution will be discussed in the next section \ref{applications}.
 
 \subsection{Interpretation}
 $\bullet$ As explained above the nature of the effect   (\ref{RR1})  is    the same as  the  conventional Unruh effect~\cite{Unruh:1976db} discussed in section \ref{Rindler} when the  Minkowski  vacuum $ \left| 0_M \right>$ is restricted to the Rindler wedge  with no   access to the entire space time. A  pure quantum state $ \left| 0_M \right>$ becomes a thermo mixed state as a result of this quantum restriction.  The result (\ref{RR1}), by definition, implies that
   the states which were unphysical (in Minkowski space) lead to physically observable phenomena, though it can not be interpreted in terms of pure states of individual particles, see Appendix A for details.  
The effect is obviously  IR in nature, and it is basically due to the presence of the horizon which itself dynamically emerges as a result of strong interactions as advocated in refs.
~\cite{Kharzeev:2005iz,Kharzeev:2005qg,Kharzeev:2006zm,Kharzeev:2006aj,Castorina:2007eb,Satz:2008zza,Becattini:2008tx,Castorina:2008gf}. 
   \\
 $\bullet$ One can explicitly see why the cancellation   of unphysical degrees of freedom $\phi_1$ and $\phi_2$ in Minkowski space fail  to hold  for the accelerating  Rindler observer (\ref{RR1}).  
  The   selection of  the physical Hilbert subspace  (\ref{gb}) is based on the properties of the operator which selects   positive -frequency modes with respect to Minkowski time $t$. At the same time the Rindler observer selects the physical Hilbert space (\ref{gb-R}) by using positive -frequency modes    with respect to observer's proper  time $\eta$. These two sets are obviously not equivalent, as e.g. they represent a mixture of positive and negative frequencies modes defined in R- and L- Rindler wedges.  At the same time, the Rindler  observers  do not ever have access to the entire space time. Therefore, from the Rindler's view point the cancellation in Minkowski space can be only achieved if one uses both sets (L and R). Of course, using the both sets would contradict to  the basic principles as the R-Rindler observer does not have access to the L wedge even  for arbitrary small  acceleration parameter $a$. \\
  $\bullet$ This is not the first time when unphysical (in Minkowski space) ghost
contributes to a physically observable quantity.   The first example is   the famous resolution of the $U(1)_A$ problem in QCD, see section \ref{Minkowski}. As long as we work in Minkowski spacetime the two constructions (based on the Veneziano ghost~\cite{ven,vendiv} 
and on the Witten's subtraction constant~\cite{witten}) are perfectly equivalent as the subsidiary condition~(\ref{gb})  ensures that the ghost degrees of freedom are decoupled from the physical Hilbert subspace, leaving both schemes with the identical physical spectrum.  
\exclude{ In an accelerating frame, on the other hand, we argued that the ``would be'' unphysical ghost  can produce  a positive physical contribution to the energy-momentum tensor (\ref{RR1}).  The question arises naturally: where is the corresponding physics hidden in the language of Witten when the ghost degrees of freedom do not even enter the system? We refer to section 3.3 of   paper\cite{UZ} where this question has been  elaborated. Here we  just want to mention that the corresponding physics does not go away, but rather, it is  hidden in the boundary conditions and necessity to sum over all topological  sectors of the theory. In accelerating frame  the Witten's subtraction constant 
becomes a ``subtraction function" which is difficult to compute. }
In different words,
these unphysical degrees of freedom do not contribute to {\it absorptive parts}, but only   to the {\it real parts} of the correlation functions. In Minkowski space such a contribution  is normally  represented by a ``subtraction constant",
while in a time dependent background this subtraction constant 
becomes a ``subtraction function" which depends on acceleration. 
We advocate the ghost- based technique  to account for this physics because the corresponding   description   can be easily generalized for accelerating background, while a similar  generalization  (without the ghost, but with explicit 
accounting for the infrared behaviour at the boundaries/horizon) is unknown and  likely to be much more technically complicated. In different words, {\it the description in terms of the ghost is a matter of convenience in order to  effectively account for  the boundary/horizon effects  in topologically nontrivial sectors of the theory.} \\
$\bullet$ One should emphasize  that 
the    Veneziano ghost   we are dealing with in this paper 
is  very different from all other ghosts, including the conventional Fadeev Popov ghosts. The Veneziano ghost is not an asymptotic state, it does not propagate, it does not contribute to the absorptive parts of the correlation functions, 
as explained in Appendix A, though,  it does fluctuate and does contribute to the energy through the boundary/horizon conditions (similar to the Casimir effect).  
The spectrum of these fluctuations is very different from conventional Fadeev Popov ghosts (when momenta could be arbitrary large in order to cancel the corresponding unphysical polarizations of the gauge fields). A typical frequency of the Veneziano ghost is determined by the horizon scale
$\omega\sim a$, while the higher frequency modes $\omega\gg a$ are exponentially suppressed. \\
$\bullet$ The unique feature of   the Veneziano  ghosts is due to its  close connection to the  topological properties of the theory. Indeed,  the topological density operator $q $   is explicitly expressed in terms of the Veneziano ghost $\phi_1$  as follows, $q\sim\Box\Phi\sim \left(  \Box\hat\phi -  \Box\phi_1 \right)$ such that the contact term (representing  the real, not absorptive part of  the topological susceptibility)  $\sim \delta^4 (x) $ in eq. (\ref{top1}) is saturated by the ghost.   One should also note that  the appearance of the ghost degree of freedom  in the formalism can be traced  from   the induced  $ q^2  $ term (\ref{lag})
which contains $\Phi\Box^2\Phi$ operator (\ref{lag1}).  As is known the $ \Box^2  $ operator can be always re-written in terms of a degree of freedom with a negative kinetic term. This explains the origin and  uniqueness of the Veneziano ghost and its relation to topological features of the theory.  A number of very nontrivial properties    of this ghost   which are discussed  in this paper are intimately related to its  topological nature. \\
     \section{Observations of  the   ${\cal P}$ and ${\cal CP}$ odd fluctuations at RHIC}\label{applications}
     The goal of this   section is to apply our previous formal analysis to the very concrete subject: we want to interpret the recent  RHIC experimental results~\cite{Voloshin:2004vk,Selyuzhenkov:2005xa,Voloshin:2008jx,Abelev:2009uh,Abelev:2009tx}   as  violation of  local
     ${\cal P}$ and ${\cal CP}$ symmetries in QCD.
      The key point of all our previous discussions  can be formulated in one line:
      QCD supports the  topologically nontrivial unique vacuum fluctuations (Veneziano ghost) in the accelerating system. The fluctuations are    IR in nature, sensitive to the horizon scale $\lambda\geq a^{-1}$ for arbitrary small $a$,   they do not propagate, do not contribute to the absorptive parts of the correlation functions, but they do contribute to the real portion of the correlation functions. 
      Their IR nature is protected by topology: they remain gapless even in the presence of the strong confined forces.
      Such a property  is in huge contrast with   conventional fast quark and gluon fluctuations which have a sharp cutoff at  wavelengths
      $\lambda\sim \Lambda_{QCD}^{-1}$.
      These topological fluctuations have $0^{-+}$ quantum numbers, and in all respects very similar to the induced, slowly fluctuating  $\theta_{ind}$ discussed in section \ref{P}  with $\dot{\theta}_{ind}\sim a $. We know about the existence of the Veneziano ghost from the resolution of the $U(1)_A$ problem when it saturates the contact term in the topological susceptibility. In the accelerating frame this contact term becomes a ``subtraction function" and the corresponding  topological  fluctuations  lead  yet to another observable phenomena as we shall argue below.
      \subsection{The basic picture}
     In what follows we assume that  $ 1~ {\text{GeV}} \gg a\gg m_q$ such that we can separate         the topological fluctuations with very large 
     wave lengths $\lambda\geq a^{-1}$  which carry  $0^{-+}$ quantum numbers  from conventional fluctuations of quarks and gluons  with typical $\lambda\sim 
      1~ {\text{GeV}}^{-1}$.   We also  neglect the interacting term $\sim m_q$ in eq. (\ref{lagKS}) such that our consideration of free fields 
      in accelerating frame leading to
      (\ref{RR1}) is justified.

      Our basic picture in this regime can be formulated as follows. The conventional quark and gluon fluctuations with typical $\lambda\sim 
      1~ {\text{GeV}}^{-1}$ are propagating in the environment of very slow topological fluctuations with wave lengths $\lambda\geq a^{-1}$.
      These slow  topological fluctuations can be thought as ${\cal P}$ and ${\cal CP}$ odd environment for the fast conventional fluctuations with typical $\lambda\sim 1~ {\text{GeV}}^{-1}$. The fast conventional fluctuations are distributed according to the Planck formula as 
      discussed in section \ref{Rindler} and described  by eq. (\ref{Planck}). While this formula was derived for a massless scalar particle for illustration purposes, it is known that a similar  thermal distribution is  expected to hold for vector and spinor fields as well.  
      
      This picture for the fast fluctuations  is equivalent to  
      the ``new thermalization" scenario  advocated in refs.  \cite{Kharzeev:2005iz,Kharzeev:2005qg,Kharzeev:2006zm,Kharzeev:2006aj,Castorina:2007eb,Satz:2008zza,Becattini:2008tx,Castorina:2008gf} as it produces hadrons  distributed according to  the thermal law  determined by temperature
      (\ref{T}).
      The only new element of  this work is the observation that these conventional  fast  fluctuations with typical $\lambda\sim 1~ {\text{GeV}}^{-1}$ are propagating in the ${\cal P}$ and ${\cal CP}$ odd  environment  described 
      by new type of topological fluctuations with very large wave lengths $\lambda\geq a^{-1}$. We emphasize that the soft topological fluctuations 
      can not propagate to infinity by themselves as they are not asymptotic states; rather they produce the ${\cal P}$ and ${\cal CP}$ odd environment for conventional fast fluctuations which eventually will be  observed as the hadrons produced in this odd environment.  Our main conjecture is that 
      the ${\cal P}$ and ${\cal CP}$ odd fluctuations observed  at RHIC~\cite{Voloshin:2004vk,Selyuzhenkov:2005xa,Voloshin:2008jx,Abelev:2009uh,Abelev:2009tx} is a direct consequence of this odd environment.  In next subsection \ref{observations}  we present  a number  of qualitative consequences  of the entire framework  supporting  this  basic picture. Before we do so, we want to compare the  energetics of slow topological fluctuations
       with conventional fast fluctuations of quarks and gluons. 
      
      Our starting formula is the Planck spectrum  for the Veneziano ghost and its partner (\ref{RR1}) valid for $a\gg m_q$.
      Number density  of the ${\cal P}$   odd domains with size $\lambda\simeq \frac{2\pi}{\omega}$ is given by
           \be
\label{N_ghost}
 \dd N_{\omega} =   \frac{\dd^3k}{(2\pi)^3} \frac{2 }{(e^{2\pi\omega/a}-1)},
\ee
      while the total contribution to the energy associated with these soft fluctuations is  
      \be
\label{E_ghost}
 E_{ghost}\simeq \int \frac{\dd^3k}{(2\pi)^3} \frac{2\omega }{(e^{2\pi\omega/a}-1)}=\frac{\pi^2}{15}\left(\frac{a}{2\pi}\right)^4.
\ee
This number should be compared with standard   contribution to the thermal  energy associated with the fast fluctuations of $N_f$ light quark flavours and $N_c^2-1$ gluons
at temperature $T=  ({a}/{2\pi} )$, 
     \be
\label{gluons}
 E_{q+g} \simeq \frac{\pi^2}{15}  \left(\frac{a}{2\pi}\right)^4\left[ (N_c^2-1) +\frac{  7N_cN_f}{4}\right].
\ee
      Therefore, the relative energy associated with slow ghost fluctuations with $0^{+-}$ quantum numbers in comparison with conventional fast fluctuations of quarks and gluons 
      is estimated to be
      \be
      \label{kappa_1}
      \kappa\equiv \frac{ E_{ghost}}{ E_{q+g}}\sim \frac{1}{\left[ (N_c^2-1) +\frac{  7N_cN_f}{4}\right]},
      \ee
      which is numerically $\sim  0.05 $.  The effect is parametrically small at large $N_c$ and  proportional $\sim 1/N_c^2$ which is  a typical suppression  for any phenomena  related to topological fluctuations. The effects related to the ghost  obviously vanish at $a=0$ as eq. (\ref{E_ghost}) states.
      This limit corresponds to the transition to Minkowski space when the Veneziano ghost is decoupled from physical Hilbert space (\ref{gb}). The factor $  \kappa$ essentially counts number of fluctuating degrees of freedom which lead to the ${\cal P}$ and ${\cal CP}$ odd  environment. However, these degrees of freedom are not the asymptotic states, and they  do not propagate
      to infinity as explained above, and they do not contribute to the absorptive parts of any correlation functions.
      
      The information about  the ${\cal P}$ and ${\cal CP}$ odd  environment must be transferred to the conventional propagating degrees of freedom which can be observed and analysed. In the ideal world with $  a \ll 1~ {\text{GeV}} $ all strong interactions can be treated  as fast  fluctuations in slow varying 
background of the Veneziano ghost $\phi_1$ and its parter $\phi_2$ at nonzero acceleration $a \ll 1~ {\text{GeV}} $. As we mentioned above,  such background  can be thought as slowly varying effective $\theta_{ind}\neq 0$. The spectral properties of these fluctuations
are determined by eq. (\ref{N_ghost}) while its energetics is determined by eq. (\ref{E_ghost}).  Therefore, in the limit $  a \ll 1~ {\text{GeV}} $ we can apply 
 our previous knowledge about physical properties of hadrons in  $\theta_{ind}\neq 0$ background.   In particular,  all previous estimates on ${\cal P}$ and ${\cal CP}$ odd effects reviewed in section \ref{P} (including the charge separation effect as a result of  anomalous interaction of slow varying $\theta_{ind}\neq 0$ with electromagnetic field)
 remain valid in this limit when ${\cal P}$ odd domain 
 is much larger in size than conventional QCD fluctuations. Therefore, we shall not elaborate on this point in the present work.
 Instead, we concentrate below  on immediate  qualitative consequences of  the picture  developed in this work when  acceleration $``a"$ being  the key parameter of the system is parametrically small.

     \subsection{Observational consequences. The universality. }\label{observations} 
     
     $\bullet 1.$ First immediate consequence of the developed picture is the presence of ${\cal P}$ and ${\cal CP}$ odd  fluctuations in any accelerating system 
    with $a\neq 0$    including all energetic $e^+e^-, ~pp$ and $p\bar{p}$  interactions   when the thermal aspects corresponding to the universal temperature around $T_H\sim (150- 200) ~{\text MeV}$ have been already observed. According to the entire logic of our framework
     the presence the thermal aspects  in observations is resulted from the acceleration $``a"$ when the QCD vacuum structure is completely reconstructed.
     The corresponding reconstruction, among many  other things, leads to   topological   fluctuations   which play the role of  the  ${\cal P}$ and ${\cal CP}$  odd environment where hadrons are being produced. These topological fluctuations will be developed  in all energetic $e^+e^-, ~pp, ~p\bar{p}$ and heavy ion   collisions.  However,  the heavy ion collisions are unique  in comparison with other types of energetic   interactions 
    as they allow to study the dependence of these odd  effects as function of  centrality which, as we argue below, effectively corresponds to variation  of the acceleration parameter $``a"$ with centrality.\\
     $\bullet 2.$ As we reviewed in section \ref{T} the acceleration realized in nature  for $e^+e^-, ~pp, ~p\bar{p}$   (not heavy ions) collisions  is a universal number given by (\ref{string}) 
    and  close to $  a \simeq 1~ {\text{GeV}} $,  rather than a free parameter $``a"$.  A typical domain size where
    ${\cal P}$ and ${\cal CP}$ odd  fluctuations will be developed is determined by eq. (\ref{N_ghost}). The fluctuations   will be  order of $\lambda\simeq 2\pi/a\sim {\text{fm}}$ for  $  a \simeq 1~ {\text{GeV}} $ which is about  the size of  conventional fluctuations of quarks and gluons. In such  circumstances the correlations similar the ones studied  at RHIC~\cite{Voloshin:2004vk,Selyuzhenkov:2005xa,Voloshin:2008jx,Abelev:2009uh,Abelev:2009tx} are expected to be suppressed for $e^+e^-, ~pp, ~p\bar{p}$    collisions  as the formation of different  hadrons  most likely will occur in different  ${\cal P}$  odd domains rather than in one large domain.  Nevertheless, the effect being  suppressed for $e^+e^-, ~pp, ~p\bar{p}$    collisions, still does not vanish.  The intensity of the corresponding correlations 
    for $e^+e^-, ~pp, ~p\bar{p}$    collisions is predicted to have the same intensity as in heavy ion collisions for the  most central events, see below.\\
     $\bullet 3.$ This conclusion  (on suppression of the correlations for $e^+e^-, ~pp, ~p\bar{p}$ collisions)  changes drastically when we consider the heavy ion collisions.
     In this case it has been argued \cite{Castorina:2007eb} that the temperature (\ref{T}),  and therefore, the acceleration  $``a"$
     will be reduced for non-central  collisions, which for small angular momentum $J$ can be approximated as follows\cite{Castorina:2007eb},
     \beq
     \label{j}
     a(J)\simeq a_{J=0}\left(1-cJ^2\right), ~~ c>0,
     \eeq
     where $c$ is some positive constant. Reduction of the acceleration  $``a"$ will increase a  typical domain size where
    ${\cal P}$ and ${\cal CP}$ odd  fluctuations develop as eq. (\ref{N_ghost}) suggests. At the same time, the conventional fluctuations responsible for the  formation of   hadrons
     are not much affected by the reduction of 
     $``a"$ as they keep a   typical (for Minkowski space)  scale $\sim {\text{fm}}$ determined by  confinement forces rather than by  acceleration $``a"$. When the size of the ${\cal P}$ odd domain becomes sufficiently larger  than $\sim {\text{fm}}$ scale the strength of  correlations should drastically increase as quite a few particles could be  formed in the same large ${\cal P}$ odd domain. Even a slight  reduction of  $``a"$ (which corresponds to
     moving  toward the least central  collisions), may produce some drastic changes in strength of correlations as dependence on $``a"$ is exponential, see eq. (\ref{N_ghost}). Strong dependence on centrality is indeed supported by observations ~\cite{Abelev:2009tx}, though the acceleration  parameter $``a"$  is obviously  not identically the same variable as centrality defined in \cite{Voloshin:2004vk,Selyuzhenkov:2005xa,Voloshin:2008jx,Abelev:2009uh,Abelev:2009tx}.  \\
           $\bullet 4.$ Another immediate consequence of the developed picture is  that the correlations should demonstrate the universal behaviour similar to the universality discussed in section \ref{T} as the source for the both effects (`` new thermalization" scenario and  ${\cal{P}}$  and ${\cal{CP}}$ odd
          effects in QCD)  is the same as argued in this paper. In particular, the effect should not depend 
       on energy of colliding ions. Indeed, the size of the ${\cal P}$ odd domains as well as the spectrum of the formed particles is determined exclusively by the acceleration $``a"$ and should not depend on energy of colliding ions. Such independence on energy is indeed supported by observations        where correlations measured in ${\text{Au+Au }} $ and ${\text{Cu+Cu }} $ collisions at $\sqrt{s_{NN}}= 62~ {\text{GeV}} $ and  $\sqrt{s_{NN}}= 200~ {\text{GeV}} $  
       are almost identical and  independent on energy~\cite{Abelev:2009tx}. These  similarities in  behaviour of the correlations is definitely  a strong argument supporting entire framework based on universality and common origin of both effects as formulated in introductory section \ref{motivation}.
       Based on this universality we predict that the corresponding correlations at the LHC energies would demonstrate a similar strength and a similar
       features found at RHIC.  \\
  $\bullet 5.$     One should emphasize that the universal features as formulated above are  related exclusively to the portion of the ``apparent thermalization"  of the system as a result of   acceleration $``a"$. Those aspects    are expected to be universal for all high energy collisions: from  $e^+e^- $ to $AA$. In case of heavy ion collisions however,  in addition to those ``apparent thermalization"  aspects   there are very real  thermodynamical features 
  of the system resulting from the conventional collisions which are normally described using the  hydrodynamical equations. This conventional
  ``hydro" portion of the dynamics, of course is not universal. This portion,  for example, is not present in $pp$ collisions, and  must be subtracted from analysis when comparison of $AA$ with $pp$ collisions is made in order to test the  universality conjecture.      \\
  \exclude{
       \begin{figure}[t]
\begin{center}
 \includegraphics[width = 0.4\textwidth]{200GeV.pdf}
  \includegraphics[width = 0.4\textwidth]{62GeV.pdf}
 \caption{\label{RHIC}
Data for $\sqrt{s_{NN}}=200\text{GeV}$ and $\sqrt{s_{NN}}=62\text{GeV}$  for Au+Au and Cu+Cu collisions  (adapted  from~\cite{Abelev:2009tx}). 
The plots demonstrate the universality in behaviour, see item 4 in the text.}
\end{center}
\end{figure}
}
     $\bullet 6.$ The arguments presented above on universal behaviour   
      do not   explicitly    depend on the strength of the magnetic field which is a key player in CME, see eq. (\ref{J}). 
      This   is a consequence  of  the   same universal behaviour  discussed above.   The direction of  $\vec{B}$  (or angular momentum $\vec{L}$)  does play a   role    of    selecting  the reaction plane, while  the absolute value of $|\vec{B}|$ is less important  parameter in our arguments. 
      The corresponding $|\vec{B}|$- dependence    is implicitly hidden
      in the magnitude of acceleration  parameter $``a"$ which is a function of many other things, including centrality, $|\vec{B}|$, etc. Therefore, one should not expect a strong dependence 
      of the effect on charges $Z_i$ of colliding ions which would lead to very different magnetic  fields  $|\vec{B}|$ for  ${\text{Au+Au }} $ and ${\text{Cu+Cu }} $ collisions.
      Observed  similarity in behaviour for  ${\text{Au+Au }} $ and ${\text{Cu+Cu }} $   is another manifestation  of  universality discussed in item 4 above. A relatively mild   $Z_i $ dependence  of the effect  is indeed   consistent with observations ~\cite{Abelev:2009tx}.\\
      $\bullet 7. $ The arguments presented above on universality of the   correlation strength 
     do not depend  on transverse momenta $k_{\perp}^2$, even for relatively  large $k_{\perp}> 1~ {\text{GeV}} $. This   is a consequence  of  the   same universal behaviour  discussed in items 4 and 5. Indeed, the entire picture described above, assumes 
     that  all  hadrons  are formed with    $k_{\perp}$  determined by the  thermal distribution in the   ${\cal{P}}$  odd background (\ref{N_ghost}). 
     The spectrum of  both: slow and fast fluctuations is the result of preparation of the vacuum  state $|0_M\ra$ in the accelerating frame even before the collision develops
     as explained above and  expressed by eq. (\ref{squeezed}). Therefore, 
   one should not expect  strong  dependence on $|k_{\perp, \alpha}-k_{\perp,\beta}| $ in the correlations for particles $\alpha$ and $\beta$ 
   even for large $|k_{\perp, \alpha}-k_{\perp,\beta}| \geq 1~ {\text{GeV}}  $ as the corresponding distributions are not much affected by ${\cal{P}}$  odd fluctuations (\ref{N_ghost}).
      This consequence  of the universality  is also  consistent with observations ~\cite{Abelev:2009tx} where it is found that the correlation depends very weakly on $|k_{\perp, \alpha}-k_{\perp,\beta}| $.\\
    $\bullet 8.$      A     picture outlined   above assumes  an ideal world with $  a  \ll  1~ {\text{GeV}} $ when a very large  ${\cal{P}}$  odd domain is formed
     with  size $\sim (2\pi)/a $   which is much larger than   conventional QCD fluctuations  with  typical sizes  $\sim {\text{fm}}$. In reality, one should expect some deviations from this universal behaviour due to  a number of complications
     in the real (rather than ideal) world, e.g. finite size of the system.  In particular, 
      the strength of the  correlations is expected to  increase when   $|k_{\perp, \alpha}+k_{\perp,\beta}| $ increases for finite (rather than very large)    ${\cal P}$ odd domain. This is   
    due to  the fact that the probability to form two $  {\text{fm}}-$ size particles within one {\it finite size domain}  is larger if the particles have    {\it smaller}   sizes, and  correspondingly larger    $|k_{\perp, \alpha}+k_{\perp,\beta}| $.  This deviation from the universality  apparently  consistent with observations ~\cite{Abelev:2009tx} where it is found that the correlation in fact increases
      with  $|k_{\perp, \alpha}+k_{\perp,\beta}| $ even for relatively  large $k_{\perp}> 1~ {\text{GeV}} $. Such a behaviour  naively contradicts to  a conventional intuition that all non-perturbative effects must be suppressed for large $k_{\perp}> 1~ {\text{GeV}} $ but in fact it has a simple and natural explanation within our framework as suggested above. Another manifestation of the same finite size effect would be a   sharp 
     cutoff in the correlations when the centrality  continues to  increase. This   pure geometrical effect 
     occurs when the available overlapping portion of the colliding    nucleus and the size of a ${\cal P}$ odd domain become approximately equal in sizes. The observation of the corresponding  peak  in strength of the correlations gives a precise  experimental tool to measure  the maximal effective size of  ${\cal P}$ odd domains   for a given system. 
     This feature, of course, can  not have an universal description as it is related to the finite size effects, and therefore,
     is sensitive to specific properties  of colliding nucleus.  
   
     To conclude: The qualitative consequences which follow from the picture outlined above apparently consistent with all presently available data. 
     In reality  $``a"$ is not a small number, and  the  size of  ${\cal P}$ odd domain is not very large even for non-central collisions. The finite size effects  and other non-universal features may lead to some corrections from the universal picture 
    as we mentioned in item  8 above. Moreover, those  corrections themselves are not expected to follow the universal behaviour. 
     Much work needs  to be done before a qualitative picture sketched  above  becomes a quantitative description of  the       correlations observed at RHIC ~\cite{Voloshin:2004vk,Selyuzhenkov:2005xa,Voloshin:2008jx,Abelev:2009uh,Abelev:2009tx}.

      \section{Conclusion. Future Directions.}\label{conclusion} 
 In this paper we adopt  the approach suggested in refs.~\cite{Kharzeev:2005iz,Kharzeev:2005qg,Kharzeev:2006zm,Kharzeev:2006aj,Castorina:2007eb,Satz:2008zza,Becattini:2008tx,Castorina:2008gf} and treat the universality 
 observed in all high energy collisions as a result of the Unruh radiation characterized by a single parameter $``a"$.
 The problem of computation the acceleration  $``a"$  is not addressed in the present paper. 
 It is  is obviously very hard problem of strongly interacting QCD.   Instead, we study the topological fluctuations (represented in our framework by the Veneziano ghost) in the given accelerating background $``a"$.  Such a treatment  is a consistent description in large $N_c$ limit when the influence of these degrees of freedom on acceleration $``a"$ itself  is negligible. 
 There is a number of immediate consequences from this picture relevant for analysis of the  correlations 
 observed at RHIC~\cite{Voloshin:2004vk,Selyuzhenkov:2005xa,Voloshin:2008jx,Abelev:2009uh,Abelev:2009tx} and which were   outlined in section \ref{observations}. We formulate the universal properties  for  the ${\cal P}$ odd  effects,
 similar to well-known universal thermal behaviour  of the spectrum studied   in all high energy collisions.   We observe that our predictions are consistent with all presently available data. 
 
 We formulate below some possible directions for the future study which may confirm or rule out this entire framework.\\
   $\bullet$ It would be very desirable (if not crucial) to understand the connection between  the acceleration parameter $``a"$ which is the key element of the present paper and familiar   parameters such as   centrality, initial energy and  the charges of the colliding ions in realistic (rather than ideal) world. Such knowledge would allow us to (quantitatively) test the basic conjecture on  universality  of the ${\cal P}$ odd  effects formulated in section \ref{observations}.\\
  $\bullet$ It would be very interesting to study  the ${\cal P}$ odd  correlations for other high energy collisions, beyond the heavy ion systems. As we mentioned in the text the ${\cal P}$ odd domains are also produced in 
  $e^+e^-$, $pp$ and $p\bar{p}$  systems, though  the size of the produced ${\cal P}$ odd domain would be  quite small as it is determined by $2\pi/a$ with $a$ given  by  eq. (\ref{string}). If one uses the same correlation 
   \cite{Voloshin:2004vk}  which has been used  for analysis of  heavy ion system, the universality arguments suggest that the intensity of the correlations   in $e^+e^-$, $pp$ and $p\bar{p}$  systems would be exactly the same as measured in heavy ions in most central  collisions when the effective acceleration $``a"$ is determined by the same expression for
   all systems (\ref{string}).\\
  $\bullet$ The Veneziano ghost  which is the subject of this paper may in fact lead to another IR effect demonstrating the sensitivity 
  to the boundaries (in addition  to sensitivity to  the horizon  scale studied in this work). Specifically,  the Veneziano ghost which is protected by topology and which 
    saturating  the subtraction term in the topological susceptibility,  may lead to the Casimir type effects  as   argued 
   in~\cite{our4d} though    no massless physical degrees of freedom are present in the system. 
  This effect can be exactly computed in  2d QED  which  is known to be the model with a single massive degree of freedom ~\cite{KS}.
  Still, the Casimir like effect is  present in this system~ \cite{toy}.  The Casimir type effects 
  in 4d QCD appear to be present  
  on the lattice where the power like behaviour $(1/L)^{\alpha}$ as a function of the total   lattice size $L$  has been observed in measurements of the topological susceptibility~\cite{Gubarev:2005jm}. Such a behaviour is in huge contrast with  exponential $\exp(-L)$ decay  law which   one normally expects for any  theories with  massive degrees of freedom\footnote{ I am thankful to Misha Polikarpov who brought the paper \cite{Gubarev:2005jm} to my attention.}.\\
   $\bullet$ The obtained results may have some   profound consequences for our understanding of physics at the largest possible scales in our universe as a result of dynamics of the same  (protected by topology) Veneziano ghost.
   Namely, the Casimir type effects 
  in 4d QCD may be observable using the CMB analysis  as suggested in ref. \cite{cmbt}. 
   \exclude{In particular, as we argued in section \ref{CMB}, the  ${\cal P}$ odd  effects studied at RHIC  and analyzed in CMB sky may have a common origin though at drastically different scales. In this relation the universe corresponds to a single large  ${\cal P}$ odd domain produced in heavy ion collisions   at the smallest possible  centralities when a typical  size of the ${\cal P}$ odd domain $\sim (2\pi/ a)$ greatly exceeds a typical size of conventional QCD fluctuations $\sim {\text{fm}}$.} 
    \\
  $\bullet$  Another manifestation of the same physics is as follows.  
 The dark energy observed  in our universe might be  a result of mismatch between the vacuum energy computed in slowly expanding universe with the expansion rate $H$  and the one which is computed in flat Minkowski space. If true, the difference between two metrics would lead to an estimate $\Delta E_{vac}\sim H\Lambda_{QCD}^3\sim (10^{-3} {\text eV})^4$ which is amazingly close to the observed value today\cite{UZ}.    As explained in the text   the typical wavelengths $\lambda$    associated with this energy density are of the order of the inverse Hubble parameter, $\lambda \sim (2\pi/H)\sim 10\textrm{~Gyr}$, and therefore, these modes  do not clump on distances smaller than $H^{-1}$, in contrast with all other types of matter. This    makes these fluctuations to be the  perfect dark energy candidates~\cite{UZ}.\\
  $\bullet$ Also, a nature of the   magnetic field in the universe with characteristic intensity of around a few $\mu$G   correlated on very large scales and observed today  is still unknown.
   One can argue that the very same Veneziano ghost  which is the subject of the present work may in fact induce
 the large scale magnetic field with correlation length $ \sim (2\pi/H) $ as a result of anomalous interaction  similar to the one which leads to the charge separation   and chiral magnetic effects (\ref{J}). In the case of heavy ions the    correlation length for charge separation effect and CME is determined by  the size of the ${\cal P}$ odd  domain $\sim (2\pi/ a)$, see   section \ref{P},
 while in case of expanding universe  the correlation length $ \sim (2\pi/H) $.  More than that,  the corresponding induced magnetic field in the universe is expected to be helical (i.e. $\int d^3x \vec{A}\cdot \vec{B}\neq 0$) and would naturally have the  intensity  $ B \simeq \frac{\alpha}{2\pi} \sqrt{H \Lqcd^3}\sim$ nG, which  by simple adiabatic compression during the structure formation epoch, could explain the field observed today at all scales, from galaxies to superclusters~\cite{magF}.\\
$\bullet$ Finally, the cosmological observations on the largest scales exhibit a solid record of unexpected anomalies and alignments, apparently pointing towards a large scale violation of statistical isotropy.  These include a variety of CMB measurements, as well as alignments and correlations of quasar polarization vectors  over huge distances of order of 1 Gpc.   The only comment I would like to make here that such anomalies may in fact be originated 
from   the same fundamental topological ${\cal P}$ odd fluctuations  studied  in this work, see~\cite{Urban:2010wa} for the details.

To conclude: the development of the early Universe is a remarkable laboratory for the study of most nontrivial properties of particle physics such as ${\cal P}$ odd effects on the scale  of the entire universe. What is more remarkable is the fact that the very same  phenomena   can be, in principle, experimentally tested in heavy ion collisions, where a similar  environment  can be achieved.  

     \section{Acknowledgements}
 I am    thankful to  Dima Kharzeev, Larry McLerran,  Valery Rubakov, Edward Shuryak, and also  Berndt Mueller and  Andrey Leonidov and other participants of the   workshops ``${\cal P}$ - and ${\cal CP}$ -odd Effects in Hot and Dense Matter" at Brookhaven, April 2010  
 and  ``the first heavy ion collisions at the LHC"  at CERN, August 2010,   where this work has been presented, 
  for useful and stimulating discussions. I am also  thankful to    James Bjorken for many hours of discussions during his visit to  Vancouver in October 2010.
 This research was supported in part by the Natural Sciences and Engineering Research Council of Canada.

 \appendix
  \section{The Veneziano  ghost is not an asymptotic state. }
  
  The main goal of this Appendix is to argue that while the Veneziano ghost leads to a number of profound effects in 
  Minkowski space (the resolution of the $U(1)_A$ problem) as well as in the accelerating frame (present work), it nevertheless is not a conventional propagating asymptotic state. In different words, it does not contribute to absorptive parts of any correlation functions. However, it does contribute to the real parts of the correlation functions. 
    In Minkowski space such kind of contributions   normally treated as the subtraction constants. In the present case of the accelerating frame, the corresponding ``subtraction constant" becomes a ``subtraction function" which depends on acceleration and which is sensitive to the global properties of the space-time. 
  \exclude{  Is it a real physical effect? One should remind the reader that a concern of the ``reality"  of the Unruh radiation was  unsettled until the paper \cite{Unruh:1983ms} appeared, see also~\cite{Birrell:1982ix}. The paper was 
  specifically devoted to the ``reality" issue. To be more specific, the authors of ref.  \cite{Unruh:1983ms} consider a simple particle detector model to demonstrate that 
  the radiation is a real physical phenomenon resolving  a number of paradoxes related to causality and energy conservation. An important for the present work result of ref.  \cite{Unruh:1983ms} is as follows: the absorption of a Rindler particle corresponds to emission of a Minkowski particle without violation causality and energy conservation. Now we want to repeat a similar analysis to see if any physical  radiation really occurs in our case when the  system is described by two fields $\phi_1, \phi_2$ with opposite commutation relations (\ref{lag}), (\ref{comm2}), (\ref{comm1})  instead of a single physical massless field in ref. \cite{Unruh:1983ms}. }
  
  The crucial observation for future analysis is as follows:
  the fields $\phi_1, \phi_2$ which are originated from unphysical (in Minkowski space) degrees of freedom can couple to  other fields only through a combination $(\phi_1-\phi_2)$ as a consequence  of the original gauge invariance 
  (\ref{lagKS}). Precisely this property along with Gubta-Bleuler auxiliary condition (\ref{gb}) provides the decoupling of physical degrees of freedom from unphysical combination $(\phi_2 - \phi_1)$
  as discussed in great details in \cite{KS}. 
  
  We follow analysis of ref.  \cite{Unruh:1983ms} to study the propagating features of fields in accelerating  
  frame. To achieve this goal we   replace  a single physical field $\Phi$ from ref. \cite{Unruh:1983ms} by specific combination $(\phi_2 - \phi_1)$ fields for our system (\ref{lagKS}).  It leads to some drastic consequences as instead of conventional expectation values such as 
  $<0| a_k ... a^{\dagger}_{k'}|0>\neq 0$ from ref. \cite{Unruh:1983ms} we would get $<0| (a_k-b_k)... (a^{\dagger}_{k'}-b^{\dagger}_{k'})|0>= 0$. The relevant  matrix elements vanish as a result   of the corresponding commutation relation $[(a^{\dagger}_{k'}-b^{\dagger}_{k'}), (a_k-b_k)]=0$. Furthermore,  as $[\mathrm{H}, (a_k-b_k)]=(a_k-b_k)$ the structure 
  $(a_k-b_k)$ is preserved such that $a_k$ and $b_k$ never appear separately. Based on this observation, 
    one can argue that the same property holds  for any other operators which constructed from the combination $(\phi_2 - \phi_1)$. In different words, no actual radiation of real particle occurs in our case in contrast with real Unruh radiation \cite{Unruh:1983ms}, i.e  the {\it absorptive parts}  of relevant correlation functions always  vanish. 
  Therefore, there are some  fluctuating degrees of freedom 
  in the system observed by a Rindler observer  without radiation of any real particles.  In many respects, this feature  is similar to the Casimir effect  though spectral density distribution  describing the fluctuations of the vacuum energy has a nontrivial $\omega$ dependence in contrast with what happens in the Casimir effect.

  Another way to arrive to the same conclusion is to consider    the particle detector moving along the world line described by some function $x^{\mu}(\tau)$ where $\tau$ is the detector's proper time. In the case for the Rindler space the corresponding $\tau$ is identified with $\eta$ defined by formula (\ref{eta}). As is known, the corresponding analysis
in the lowest order approximation   is reduced to study of the positive frequency Wightman Green function defined as 
\be
\label{wightman}
D^+ (x, x')= \< 0|\Phi (x), \Phi (x') |0\> , 
\ee
while the transition probability per unit proper time is proportional to
its Fourier transform, 
\be
\label{fourier}
\sim \int^{+\infty}_{-\infty} d(\Delta \tau)e^{-i\omega\Delta \tau} D^+ (\Delta\tau) 
\ee
where we use notations from ~\cite{Birrell:1982ix}.   
 In case of inertial trajectory for massless scalar field $\Phi$  the positive frequency Wightman Green function
 is given by 
 \be
D^+ (\Delta\tau')= -\frac{1}{4\pi^2} \frac{1}{(\Delta\tau -i\epsilon)^2}
\ee
and the corresponding Fourier transform (\ref{fourier}) obviously vanishes. No particles are detected as expected.
In case if the detector accelerates uniformly with acceleration $a$   the corresponding Green's function is given by~\cite{Birrell:1982ix}
 \be
D^+ (\Delta\tau')= -\frac{1}{4\pi^2}\sum_k \frac{1}{ \left(\Delta\tau   -i2\epsilon +2i\pi\frac{k}{a}\right)^2}.
\ee
As there are infinite number of poles in the lower -half plane at $\Delta\tau =- 2i\pi\frac{k}{a}$ for positive $k$ the corresponding Fourier transform (\ref{fourier}) leads to the   known   result $\sim  \omega[\exp(2\pi\omega/a)-1]^{-1}$.

In our case the detector- field interaction is described by the combination $(\phi_1-\phi_2)$ rather by a single field $\Phi$ discussed above. Therefore, the relevant response function in our case
is described by the positive frequency   Green's function defined as 
\be
\label{ghost}
 \sim \< 0|\Big(\phi_1(x)-\phi_2(x)\Big),\Big(\phi_1(x')-\phi_2(x') \Big) |0\> , 
\ee
which replaces eq. (\ref{wightman}). One can easily see that this  Green's function given by eq. (\ref{ghost}) identically vanishes as the consequence of the opposite  signs in commutation relations  describing $\phi_1$ and $\phi_2$ fields, in complete agreement with the arguments  presented above. Therefore, the Rindler observer will see an extra energy (\ref{RR1}) without detecting any physical particles. One can rephrase the same statement by saying that the ghost and its partner do not contribute to the absorptive part of the Green's function, but  do contribute    to its  real part. 
In Minkowski space the contribution to the real part of the topological susceptibility is nothing but the well known subtraction constant which has been precisely studied and measured on the lattice, see   Fig.\ref{chi-lattice}. 
The ghost- based technique we advocate in this paper is well suited  to study  the corresponding  physics in accelerating or time dependent background.

\exclude{
This picture is based, of course, on the standard treatment of gravity as a background field. Such an approximation is justified as long as the produced 
effect  is much smaller than the background field itself. Otherwise, the feedback reaction must be considered. The corresponding analysis, however,  is beyond 
  the scope of this work,  and shall not be discussed here. 
}

\end{document}